%% file: artigoum_arxiv.tex
\documentclass[10pt]{article}
\usepackage{amsmath,amsfonts,amssymb,longtable,placeins,graphicx,lscape}
\usepackage{setspace}                   
\usepackage[authoryear]{natbib} 
\usepackage[letterpaper,top=3cm, bottom=3cm, left=3cm, right=3cm]{geometry}
\usepackage{multirow}
\usepackage{rotating}  

\newcommand {\dps}{\displaystyle}
\newcommand {\rg}{\rightarrow}

\begin{document}


\title{Wavelet Functional Data Analysis for FANOVA Models under Dependent Errors}

\author{Airton Kist $^{(1)}$, Alu\'{\i}sio Pinheiro$^{(2)}$\footnote{corresponding author: pinheiro@ime.unicamp.br}\\ $^{(1)}$
University of Ponta Grossa, Brazil, $^{(2)}$ University of Campinas, Brazil}
\date{}
\maketitle \abstract{We extend the wavelet tests for fixed effects FANOVA models with iid errors, proposed in
Abramovich et al, 2004 to FANOVA models with dependent errors and provide an iterative Cochrane-Orcutt type procedure to
estimate the parameters and the functional. The function is estimated through a nonlinear wavelet estimator.
Nonparametric tests based on the optimal performance of nonlinear wavelet estimators are also proposed. The method is
illustrated on real data sets and in simulated studies. The simulation also addresses the test performance under
realistic sample sizes.


}

\section{The model}

Consider the Ornstein-Uhlenbeck diffusion process as follows:
\begin{equation}\label{OU-yt}
dy(t)=\rho_c y(t)dt+\sigma dW(t);\quad y_0=b,\,\,t>0,
\end{equation}
where $\rho_c$ and  $\sigma>0$ are unknown fixed parameters, and $\{W(t): t\geq 0\}$ is a standard Brownian motion. The unique
solution for $\{y(t): t\geq 0\}$ in the mean square sense \citep{Arnold1974} is given by
\begin{equation}\label{OU-solyt}
y_t=e^{\rho_ct}b+\sigma\int_0^te^{\rho_c(t-s)}dW(s)=e^{\rho_ct}b+\sigma
J_{\rho_c}(t),
\end{equation}
where $J_{\rho_c}(t)=\int_0^te^{\rho_c(t-s)}dW(s)$. Note that
$J_{\rho_c}(t)\sim\mathrm{N}(0,(e^{\rho_ct}-1)/2\rho_c)$.

Suppose one is interested in estimating $\rho_c$ based on a single observed path $\{y(t)\}_{0\leq t\leq T}$. The mean
squared estimator of $\rho$ is given by\begin{equation}\label{rhoestcont}
\hat{\rho}_c=\int_0^Ty(t)dy(t)\Big/\int_0^Ty(t)dt.
\end{equation}
which is also the unrestricted maximum likelihood estimator if $b=0$.

The discrete representation of such process is given by
\begin{equation}\label{OU-ytdiscr}
y_{t,h}=e^{\rho_ch}y_{(t-1),h}+u_{t,h},~t>0,\quad y_0=b,
\end{equation}
where $u_{t,h}\sim\mathrm{N}(0,\sigma^2(e^{2\rho_ch}-1)/2\rho_c)$ and
$h$ is the sampling interval.

In this discretized version of the problem, one aims to estimate
  $\rho_h=e^{\rho_ch}$ given the observations $\{y_{t,h}\}_{t=0}^n$, where
$n=T/h$. The minimum squared error estimator, which is also the conditional (on $y_0$) maximum likelihood solution, is given by
\begin{equation*}\label{rhoesth}
\hat{\rho}_h=\sum_{t=1}^ny_{t,h}y_{(t-1),h}\Big/\sum_{t=1}^ny_{(t-1),h}^2.
\end{equation*}

\cite{Perron1991} studies the limit distribution of $n(\hat{\rho}_h-\rho_h)$ under  (\ref{OU-ytdiscr}), when  $h\rg0$ and fixed
$T$, and proves that it is identical to the limiting distribution of
$T(\hat{\rho}_c-\rho_c)$ under (\ref{OU-yt}). Suppose one is interested in estimating and testing functions $f$ in the model defined by:
\begin{eqnarray}\label{mod_ef_fixoscar1}
  dy_i(t) &=& f_i(t)dt+\varepsilon_i(t)dt,\qquad
  t\in[0,1],\,\,\,i=1,\ldots,r,
\end{eqnarray}
where $r$ is the number of curves being compared and $\{\varepsilon(t):t\geq0\}$ is a CAR(1) model. The aforementioned
CAR(1) is also called the Ornstein-Uhlenbeck process and is the only stationary solution of the PDE
\begin{eqnarray}\label{eq_difCAR1}
  d\varepsilon(t)+\alpha\varepsilon(t)dt &=& \sigma dW(t),
\end{eqnarray}
where $\alpha$ and  $\sigma$ are unknown positive parameters. Note that
\begin{eqnarray}\label{mom_cond1}
  \mathbb{E}[\varepsilon(t)|\varepsilon(0)] &=& e^{-\alpha
  t}\varepsilon(0)\\\label{mom_cond2}
  \mathrm{Var}[\varepsilon(t)|\varepsilon(0)] &=&
  \frac{\sigma^2}{2\alpha}(1-e^{-2\alpha t}).
\end{eqnarray}

Moreover,  $\{\varepsilon(t)\}$ can be written as
$$
\varepsilon(t)=e^{-\alpha
t}\varepsilon(0)+\sigma\int_0^te^{-\alpha(t-s)}dW(s)
 =e^{-\alpha t}\varepsilon(0)+\sigma J_\alpha(t)
$$
where $\dps J_\alpha(t)=\int_0^te^{-\alpha(t-s)}dW(s)$, which is distributed as  $\mathrm{N}(0,(1-e^{-2\alpha
t})/2\alpha)$. This solution is stationary,  so that $\alpha>0$ and
$\varepsilon(t)\sim\mathrm{N}(0,\sigma^2/2\alpha)\quad \forall t\geq0$.  Consequently,
\[
\begin{array}{l}
\mathbb{E}[y_i(t)|f_i(t),\sigma^2,\alpha] = f_i(t), \\
\mathrm{Var}[y_i(t)|f_i(t),\sigma^2,\alpha] =\sigma^2/(2\alpha)\mbox{ and} \\
\mathrm{Corr}[y_i(s),y_i(t)|f_i(t),\sigma^2,\alpha] =e^{-\alpha(t-s)}, \mbox{ for $s<t$}.
\end{array}
\]
\section{Estimation and Testing}

Suppose the model defined by (\ref{mod_ef_fixoscar1}). If the sample on the CAR(1) is equally spaced, i.e. such that $h=1/n$ one can write $\rho=\rho(\alpha,h)=e^{-\alpha h}$. Given
the discrete process variance $\sigma_p^2$, one can write $\sigma_p^2={\sigma^2}/(2\alpha)$, and
$\sigma_p^2(1-e^{-2\alpha h})=\sigma_p^2(1-\rho^2)$.

Following (4), we discretize the model as
\begin{equation}\label{mod_ef_fixos_discr}
    y_{i,t}=f_{i,t}+\varepsilon_{i,t},\quad i=1,\ldots r,
\end{equation}
where $\varepsilon_{i,t}=\rho\varepsilon_{i,t-1}+u_{i,t}$ are independent CAR(1) processes, $i=1,\ldots, r$, and the
$u_{i,t}\sim (0,\sigma_u^2)$ uncorrelated with $\varepsilon_{i,s}$, $s<t$ for each  $i$. The Fisher information matrix for this model is given by:
$$I(f_t,\rho,\sigma_u^2)=\left[%
\begin{array}{ccc}
 \frac{1}{\sigma_u^2}\left(1-\rho^2 + (n-1)(1-\rho)^2\right)  & 0 & 0 \\
 0  & \dps\frac{n-1+(3-n)\rho^2}{(1-\rho^2)^2} & \dps\frac{1}{\sigma_u^2(1-\rho^2)} \\
 0  & \dps\frac{1}{\sigma_u^2(1-\rho^2)} & \dps\frac{n}{2\sigma_u^4} \\
\end{array}%
\right],$$ which, being block-diagonal, justifies applying diverse methods for $f_t$ and $(\rho,\sigma_u^2)$. For instance, we estimate the former by wavelets, and the latter by ML.

The procedure can be resumed as follows:
\begin{enumerate}
    \item [(E1)] Initial solution for $\rho:\hat{\rho}\,\,\in(-1,1)$;
    \item [(E2)] Compute  $y_t-\hat{\rho} y_{t-1} = f_t-\hat{\rho}
    f_{t-1}+u_t$, i.e., $z_t = g_t+u_t$, and estimate $g$ ($g_t$) by $\hat{g}$ ($\hat{g}_t$);
    \item [(E3)] Estimate $f$ by $\hat{f}_t=\hat{g}_t+\hat{\rho}\hat{f}_{t-1}$,
    with $\hat{f}_0=y_0$;
    \item [(E4)] Estimate $\rho$ by $\hat{\rho}=\sum_{t=2}^n e_te_{t-1}/\sum_{t=2}^ne_t^2$, where $e_t=y_t-\hat{f}_t;$
    \item [(E5)] Testing convergence by subsequent estimated values of  $\rho$;
    \item [(E6)] Repete steps  (2)-(5) until convergence is attained, or the maximum number of iterations is reached.
\end{enumerate}

\vspace*{.5cm}

\noindent {\bf Remarks:} 
\begin{enumerate}
\item In step (2) one may estimate $g$ linearly on nonlinearly.
\item The variance $\sigma_u^2$ must be estimated in each iteration if non-linear wavelet estimators are used
in (2). Otherwise, one only needs it at the end of the process. In the simulation studies and in the application, we
compared MAD and STD estimates for $\sigma_u^2$ (Vidakovic, 1999 pp. 196-7).
\end{enumerate}
\vspace*{.5cm}

It is known that high-dimensional models, and functional models as well, pose a problem for the classical criteria such
as Neyman optimality. For that reason some shrinkage must be applied in order to get statistically sound solutions. In
the HANOVA setup, \cite{Fan1996} and \cite{Fan1998} present adaptive Neyman tests for high-dimensional parameters and
curves, respectively.

\cite{Abramovich2004} proposes a FANOVA model for
\begin{eqnarray}\label{mod_ef_fixosiid}
  dy_i(t) &=& f_i(t)dt+\varepsilon_i(t)dt,\qquad
  t\in[0,1],\,\,\,i=1,\ldots,r,
\end{eqnarray}
where $r$ is the number of curves being compared and $\{\varepsilon(t):t\geq0\}$ is a Brownian motion.
(\ref{mod_ef_fixosiid}) and (\ref{mod_ef_fixoscar1}) may be seen as equivalent models except for the error structure.

The FANOVA decomposition is given by
\begin{equation}\label{decomp_f_i}
f_i(t)=m_0+\mu(t)+a_i+\gamma_i(t), \quad i=1,\ldots,r; \quad
t\in[0,1]
\end{equation}
where: $m_0$ is the overall mean; $\mu(t)$ is the main effect in $t$; $a_i$ is the main effect in $i$; $\gamma_i(t)$ is
the interaction between $i$ e $t$, with the following identificability conditions
\begin{equation}\label{cond_identif}
\int_0^1\mu(t)dt=0;\,\,\,
\sum_{i=1}^ra_i=0;\,\,\,\sum_{i=1}^r\gamma_i(t)=0;\,\,\,
\int_0^1\gamma_i(t)dt=0,\,\,\, \forall \,i=1.\ldots,r;\,\,
t\in[0,1].
\end{equation}

One would be interested in hypotheses such as:
\begin{eqnarray}\label{hip_nula_mu}
&&\text{H}_0: \mu(t)\equiv0, \quad t\in[0,1];\\
\label{hip_nula_ai}
&&\text{H}_0: a_i=0, \quad i=1,\ldots,r; \text{ and}\\
\label{hip_nula_gammai} && \text{H}_0: \gamma_i(t)\equiv0,\quad
i=1,\ldots,r\quad t\in[0,1].
\end{eqnarray}

While  (\ref{hip_nula_ai}) can be treated as the usual parametric hypotheses,  (\ref{hip_nula_mu}) (\ref{hip_nula_gammai}) are
intrinsically functional. \cite{Donoho1995,Donoho1998} and \cite{Spokoiny1996} present optimal minimax rates in Besov spaces which are attained by wavelet procedures.  For a significance level $\alpha\in(0,1)$ let $\phi^\ast$ be the test defined by
\begin{equation}\label{teste_n_adaptativo}
\phi^\ast=\left\{\begin{array}{ll}
 \mathbf{1}\left\{T(j(s))>v_0(j(s))z_{1-\alpha}\right\} & \text{ if }p\geq2; \text{ or} \\
 \mathbf{1}\left\{T(j(s))+Q(j(s))>\sqrt{v_0^2(j(s))+w_0^2(j(s))}z_{1-\alpha}\right\} &
\text{ if }1\leq p<2,
\end{array}\right.
\end{equation}
where $p$, $q$, $s$ and $C$ are considered known. We refer the readers to \cite{Abramovich2004} for details. This test
is proven to be optimal in the minimax sense. The non-adaptative test also proposed by \cite{Abramovich2004}, which is
optimal as well, is given by $\phi_\eta^*=1$ if:
\begin{equation*}\label{teste_adaptativo1}
\phi_\eta^\ast=\max_{j_{\min}\leq j(s)\leq
j_\eta-1}\left\{\frac{T(j(s))+Q(j(s))}{\sqrt{v_0^2(j(s))+w_0^2(j(s))}}\right\}>
\sqrt{2\ln\ln\eta^{-2}}.
\end{equation*}

If one knows that $p\geq2$, then $\phi_\eta^*=1$ if
\begin{equation*}\label{teste_adaptativo2}
\phi_\eta^\ast=\max_{j_{\min}\leq j(s)\leq
j_\eta-1}\left\{\frac{T(j(s))}{\sqrt{v_0^2(j(s))}}\right\}>
\sqrt{2\ln\ln\eta^{-2}}.
\end{equation*}

One assumes above that the $f_i(t)$ belong to a Besov ball of radius $C>0$, in $[0,1]$, $B_s^{p,q}(C)$, where $s>1/p$ and $1\leq p, q\leq \infty$.

For the dependent error model (\ref{mod_ef_fixoscar1}), a two-step iterative procedure is employed, where in each
iteration $\rho$ and $f$ are successively estimated by a Cochrane-Orkutt type procedure, specified in (E1)-(E6), employing the
estimation (and testing) procedures proposed by \cite{Abramovich2004} for $f$.

One can then prove that reasonable error norms are minimized by this procedure. In particular, it can be proven that
the $L^2$ error norm is improved in each iteration.

\section{Simulation Studies}

\begin{figure}[!htb]
\centering
\includegraphics[width=16cm, height=8cm]{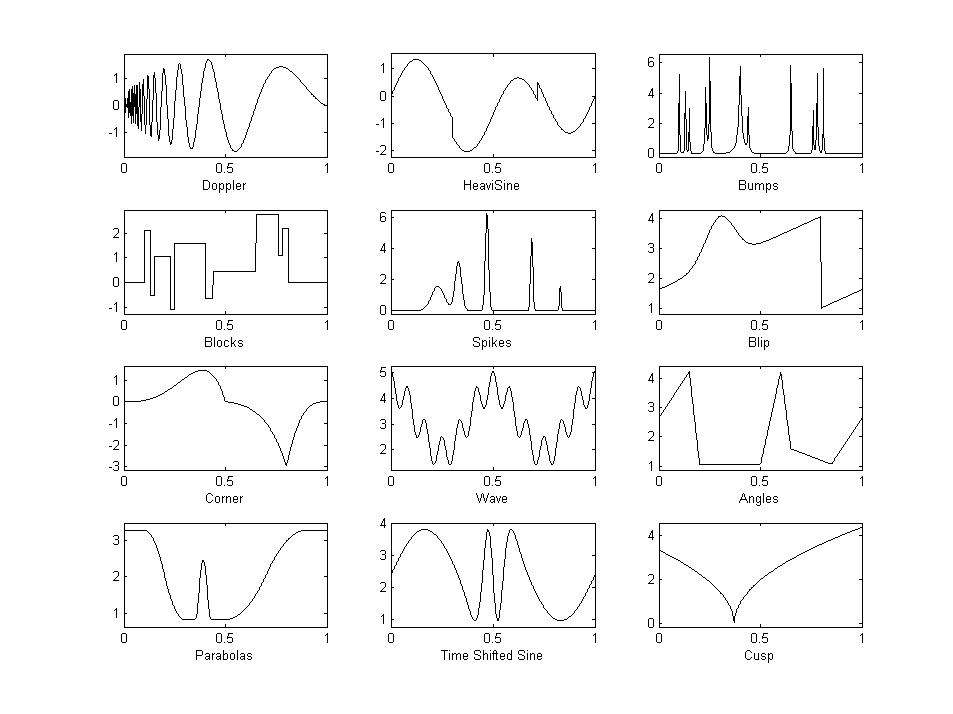}
\vspace*{-1.2cm}\caption{Test functions} \label{fc_teste}
\end{figure}

We present simulation studies to evaluate the testing procedure. Twelve classical test functions (Figure
\ref{fc_teste}), five sample sizes ($n=512;1024;2048;4096;8192$), three signal-to-noise ratios ($\mathrm{SNR}=1;3;7$),
two values of $\rho$ ($0.99$ and $0.9999$) and three wavelets bases (`db3'; `db6'; and `sym8') are considered. For each
combination, 1000 replications were taken.

Since $\alpha= -n\log\rho$ one has for $\rho=0.99$ and $n= 512; 1024; 2048; 4096; 8192$,  $\alpha= 5.14; 10.29;
20.58; 41.16; 82.33$, respectively.  For $\rho=0.9999$ , one has $\alpha =0.051; 0.102; 0.204;$  $0.409; 0.819$,
respectively.

Moreover, both linear \citep{Antoniadis1994} and non-linear \citep{Spokoiny1996, Abramovich2004} steps are implemented.
For the former projection is made in $V_5$ for $n=512$, $V_6$ for $n=1024; 2048$  and $V_7$ for $n=4096; 8192$ are employed. For
the latter, thresholding in the levels $4-7$ for $n=  512; 1024; 2048$ and  $5-8$ for $n=4096; 8192$ are performed. In the last iteration, after $\rho$ is estimated, the function is estimated by thresholding either
term-by-term \citep{Spokoiny1996,Abramovich2004}  or by blocks \citep{Cai1999,Cai2002}.

Data was generated by the model
\begin{equation}\label{modelo_sim}
    y_t=f_t+\varepsilon_t,\quad t=1,\ldots,n,
\end{equation}
where $\varepsilon_t=\rho\varepsilon_{t-1}+u_t$ is a discretized CAR(1), $\rho=e^{-\alpha/n}$,
$u_t\sim\mathrm{N}(0,(1-\rho^2){\sigma^2}/{2\alpha})$, $\varepsilon_0\sim\mathrm{N}(0,\sigma^2/2\alpha)$ and $f_t$ a
test function.

For each combination of  $\rho$ and $n$, one uses  $\sigma^2=1$ and rescales the test function to get the desired
SNR.  In each estimation procedure, $50$ randomly selected values in $(-1,1)$ are used as initial values of $\hat{\rho}$. MAD
estimates of $\sigma_u$ are employed  for the non-linear estimates. The stopping criterium was a difference on the
subsequent $\rho$ estimates smaller than $<10^{-15}$ or $250$ iterations.

A smaller preliminary simulation was performed with the single aim of assessing the real need of estimating $\rho$. Table \ref{ab} shows the performance of the estimators of $f$ as a function of the employed value of $\rho$.  1000 replications onf each combination of  $f(t)=\sin(2\pi t)$, $n=1025$,
$\mathrm{SNR}=7$ and $\rho=0.99;\,0.999;\,0.9999$ are taken. The 'db6' basis is used in a term-by-term non-linear procedure on
levels $5$ to $7$ \citep{Abramovich2004} for the final $f$ estimation. The iterations were made with thresholding from
levels $4$ to $7$. Moreover, to show the effects of ignoring the dependence on $\varepsilon$, twelve estimators are compared. Eleven fixed values for $\rho$ are considered: $-0.9,-0.7,-0.5,-0.3,-0.1,0,0.1,0.3,0.5,0.7,0.9$. The Integrated Mean Squared Error (IMSE) is used to compare the
results. We summarize the results in Table \ref{ab} through the IMSE ranks for each case ($\rho=0.99;0.999;0.9999$ or,
analogously, $\alpha=10.2915;1.0245;0.1024$). One notices that the proposed method has in general a better performance
when compared to the ad-hoc procedure based on a fixed value of $\rho$.

\begin{table}[h]
\centering \caption{Average and Median Rank on the proposed IMSE (out of 1000 replications) $f(t)=\sin(2\pi t)$,
$\mathrm{SNR}=7$, `db6'  and $n=1025$.}\label{ab}
\begin{tabular}{l|c|c|c}
\hline
Rank $\backslash\,\, \rho$  & 0.99 & 0.999 & 0.9999 \\
\hline
 Average   & 11.3820   &   11.1420   &   11.1370    \\
 Median & 11   &   12  &   12    \\
 \hline
\end{tabular}
\end{table}

We present in Table \ref{tabelaresumo} an overview for all test functions.  The lines within each column represent the number of times (out of 15) in which the titled technique or basis achieved the best performance. The Appendix has a detailed presentation
for the \emph{Doppler} test function. The details for the other test functions can be made available as Supplementary
Material.

\begin{sidewaystable}
\centering
 { \footnotesize
\caption{Overview of the Simulation Studies for Estimation of $f$, $\rho$. }\label{tabelaresumo}
\begin{tabular}{c|ccccccccccc}
\hline
Type(1) & Bias and  & \multicolumn{3}{c}{Smallest Bias} &
\multicolumn{3}{c}{Smallest MSE} &  IMSE   & \multicolumn{3}{c}{Smallest IMSE}\\
&MSE NL& `db3' & `db6' & `sym8' & `db3' & `db6' & `sym8' &NL blocks& `db3' & `db6' & `sym8'\\
\hline
Do1 & 11 & 2 & 9 & 4 & 2 & 11 & 2 & 8 & 2 & 5 & 8  \\
Do2 & 10 & 9 & 6 & 0 & 1 & 5 & 9 & 15 &1 & 5 & 9 \\
He1 & 15 & 4 & 8 & 3 & 9 & 5 & 1 & 4 & 0 & 11 & 4\\
He2 & 10 & 5 & 9 &1 & 1 & 5 & 9 & 12 & 9 & 5 & 1\\
Bu1  & 9 &  5 & 7 & 3 & 10 & 2 & 3 & 12 & 10 & 2 & 3 \\
Bu2  & 13 &  7 & 8 & 0 & 1 & 5 & 9 & 15 & 14 & 1 & 0 \\
Bk1 & 11 & 7 & 7 & 1 & 6 & 7 & 2 & 9 & 4 & 4 & 7 \\
Bk2 & 14 & 11 & 3 & 1 & 11 & 3 & 1 & 15 & 12 & 0 & 3 \\
Sp1 & 11 & 6 & 6 & 3 & 6 & 7 & 2 & 7 & 1 & 6 & 8 \\
Sp2 & 6 & 2 & 8 & 5 & 2 & 8 & 5 & 12 & 5 & 3 & 7 \\
Bp1 & 13 & 7 & 7 & 1 & 4 & 8 & 3 & 8 & 2 & 8 & 5 \\
Bp2 & 7 & 10 & 5 & 0 & 10 & 5 & 0 & 13 & 7 & 4 & 4 \\
Co1 & 15 & 6 & 3 & 6 & 3 & 7 & 5 & 4 & 13 & 0 & 2 \\
Co2 & 14 & 8 & 4 & 3 & 7 & 5 & 3 & 6 & 2 & 9 & 4 \\
Wa1 & 15 & 2 & 11 & 2 & 2 & 12 & 2 & 5 & 3 & 12 & 0 \\
Wa2 & 11 & 2 & 5 & 8 & 4 & 5 & 6 & 12 & 3 & 12 & 0 \\
An1 & 15 & 9 & 5 & 1 & 6 & 8 & 1 & 2 & 0 & 11 & 4 \\
An2 & 12 & 8 & 4 & 3 & 6 & 4 & 5 & 10 & 4 & 8 & 3 \\
Pa1 & 15 & 6 & 6 & 3 & 3 & 11 & 1 & 11 & 0 & 14 & 1 \\
Pa2 & 14 & 4 & 3 & 8 & 4 & 4 & 7 & 9 & 2 & 11 & 2 \\
Ts1 & 15 & 5 & 6 & 4 & 4 & 8 & 3 & 3 & 0 & 13 & 2 \\
Ts2 & 12 & 4 & 6 & 5 & 2 & 8 & 5 & 10 & 4 & 8 & 5 \\
Cu1 & 15 & 5 & 6 & 4 & 4 & 8 & 3 & 5 & 0 & 11 & 4 \\
Cu2 & 12 & 3 & 12 & 0 & 2 & 13 & 0 &  7 & 6 & 4 & 5 \\
\hline
\end{tabular}

\hspace*{-0.1cm} (1) Do* - \emph{Doppler}; He* - \emph{HeaviSine}; Bu* - \emph{Bumps}; Bk* - \emph{Blocks}; Sp* -
\emph{Spikes}; Bp* - \emph{Blip}; Co* - \emph{Corner}; Wa* - \emph{Wave};

\hspace*{-6.3cm} An* - \emph{Angles}; Pa* - \emph{Parabolas};
 Ts* - \emph{Time Shifted Sine}; Cu* - \emph{Cusp}.

\hspace*{-3.4cm}  For all cases, if *=1, $\rho=0.99$. If *=2, $\rho=0.9999$. NL means (nonlinear) thresholding.}
\end{sidewaystable}

Some of the simulation results were expected. For instance, for either value of $\rho$, the bias of the linear
estimator of $\rho$ generally decreases in the SNR and in the sample size. Some order discrepancies on $n$ are
observed, but the fact that the functional projection is made in a fixed $V_j$ can be taken as cause for them. A
similar behavior is seen when the non-linear estimator is employed, albeit in much milder terms. However, one should
pay special attention to the thresholding procedure. When comparing the linear and non-linear procedures, the latter is
the best performer. For $\rho=0.99$, it attains precisions of two or more decimal places against one for the former
whenever $n\geq 1024$. Analogous results are seen for  $\rho=0.9999$. In some specific situations the linear estimator
has a better performance but, since this can not be predicted in practice, the non-linear should be employed. Another
conclusion regarding the estimation of $\rho$ is that it is heavily influenced by any bias on the estimation of $f$, be
it by {\it mistaken} projection or over- and under-shrinkage. No significant effect of the initial value of $\rho$ has
been found.

The results regarding linear vs non-linear functional estimation are somewhat similar to the ones for $\rho$. There are
cases in which the linear estimator outperforms the non-linear estimator. However, there are no instance of really poor
performance by the non-linear estimator while the linear estimator fails severely for some test functions. Moreover,in
the case where the linear estimators have a better performance, the differences are small. The numerical differences
due to the basies are small but favor in general the 'db6' and 'db3' bases.

The overall recommendation is to use the term-by-term thresholding wavelet estimator for all but the last iteration.
Then after the final estimation of $\rho$, a blocking-thresholded estimator of $f$ yields the best performance.

\section{Aplication to SONDA's Environmental Data}

The data was obtained from SONDA (National System of Environmental Data, INPE-Brazil, {\it http://sonda.ccst.inpe.br}).
The SONDA network has minute-by-minute environmental data. We analyze the variables air temperateure at surface, air
relative humidity, and air pressure. The four initial months for each season, March, June, September and December, were
selected.

\begin{table}[!htb]  
  \centering
\caption{Latitude, longitude and altitude characteristics and Estimates of $\rho$ based on `db6'. $\hat{\rho}_3$,
$\hat{\rho}_6$, $\hat{\rho}_9$ e $\hat{\rho}_{12}$ are the $\rho$ estimates for March, June, September and December,
respectively.  The analysis is based upon $n=2^{15}$ data points, which corresponds to 22.75 days for  each  month. The
iterative procedure employs term-by-term thresholding from levels 5 to 8.}\label{tab1}
\begin{tabular}{cccc|cccccccccccc}
  \hline
 Station & Lat.(S) & Long.(W) & Alt.(m) & Estimator & Temperature & Humidity &  Pressure  \\
  \hline
& & & &$\hat{\rho}_3$    &  0.9996 &   0.8807 &    0.9992   \\
Bras\'{\i}lia & $15^\circ$36' & $47^\circ$42'  & 1023 & $\hat{\rho}_6$ &  0.9996 &   0.9764 &    0.9992    \\
& & & &$\hat{\rho}_9$    &  0.9995 &   0.9968 &    0.9994   \\
& & & &$\hat{\rho}_{12}$ &  0.9994 &   0.9499 &    0.9993   \\
\hline
& & & & $\hat{\rho}_3$     &  0.9988  &  0.9975  &  0.9985   \\
Ourinhos & $22^\circ$56' & $49^\circ$53'  & 446 &$\hat{\rho}_6$   &  0.9992  &  0.9978  &  0.9983   \\
& & & &$\hat{\rho}_9$     &  0.9990  &  0.9963  &  0.9985   \\
& & & &$\hat{\rho}_{12}$  &  0.9989  &  0.9932  &  0.9987   \\
\hline
& & & &$\hat{\rho}_3$     &  0.9982 &   0.9913  &  0.9995   \\
S\~ao Luiz  & $02^\circ$35' & $44^\circ$12'  & 40  &$\hat{\rho}_6$  &  0.9988 &   0.9932  &  0.9985   \\
& & & &$\hat{\rho}_9$     &  0.9005 &   0.9960  &  0.9989   \\
& & & &$\hat{\rho}_{12}$  &  0.9981 &   0.9972  &  0.9981   \\
\hline
\end{tabular}
\end{table}

The Weather stations in Bras\'{\i}lia, Ourinhos and S\~ao Luiz were chosen, given their latitude, longitude and
altitude characteristics. The integrity and the nature and consistency of the data were also important factors in the
choice of cities and variables analyzed. We consider 22.75 days for each month, which provides a total of
$2^{15}=32,768$, starting in the first hour of the first day of each selected month of 2009.

The term-to-term thresholding wavelet estimator was employed based on the `db6' basis. Five initial values for $\rho$
are employed, randomly generated from a  $U(-1,1)$.  Thresholding is performed on the levels 5 to 8 until $\rho$ is
estimated. After that, in the last iteration, only levels 8 and 9 were thresholded, as proposed by
\cite{Abramovich2004}.

Table \ref{tab1}  presents the estimates for $\rho$ for the SONDA's stations. First, we'd like to emphasize that the
final estimate does not depend on its initial value, i.e., in all cases the five initial values yield the same $\rho$
estimates. The general conclusion is that a reasonable variation is observed in the estimated values of $\rho$ which
corroborates the necessity of adjusting the data set to their effects. Moreover, high values of $\rho$ are observed.

Some specific results should be discussed. The three cities, which have very different weather conditions, do present
different behaviors in the $\rho$ values as well. For instance, Bras\'{\i}lia presents very stable values of $\rho$ for
temperature and pressure, but humidity's $\rho$ vary during the year. Ourinhos has similar behavior for the values of
$\rho$ over time and among the environmental variables. S\~ao Luiz presents reasonably stable values of $\rho$ for
humidity and pressure, but temperature in September has quite a different value of $\rho$ from the rest of the year.

\begin{figure}[!htb]
\centering
\includegraphics[width=16cm, height=8cm,height=8cm]{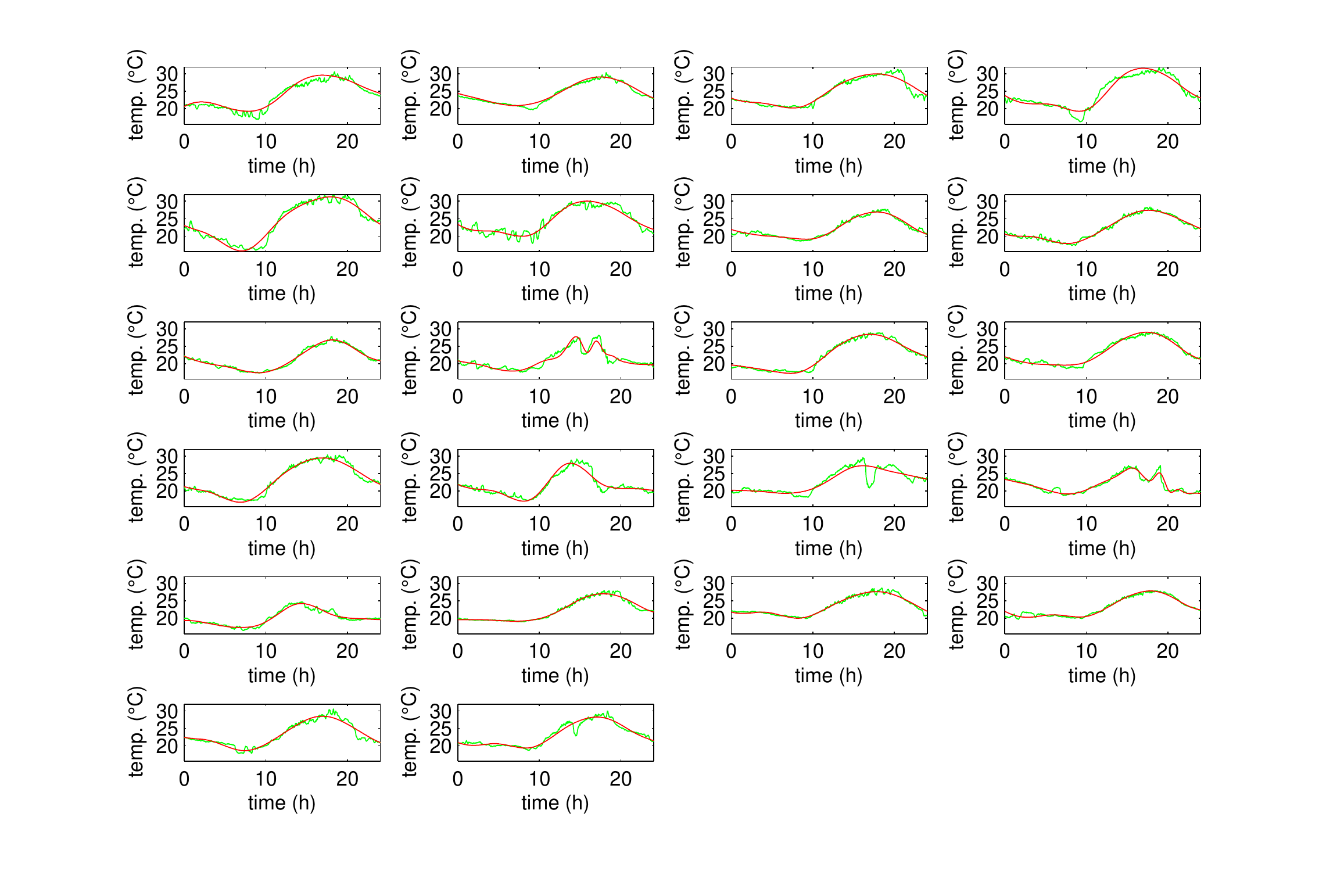}
\vspace*{-1.1cm} \caption{Bras\'{\i}lia's estimated and observed temperature curves for the month of March
2009.}\label{figtempmarBrasOS}
\end{figure}

\begin{figure}[!htb]
\centering
\includegraphics[width=16cm, height=8cm,height=8cm]{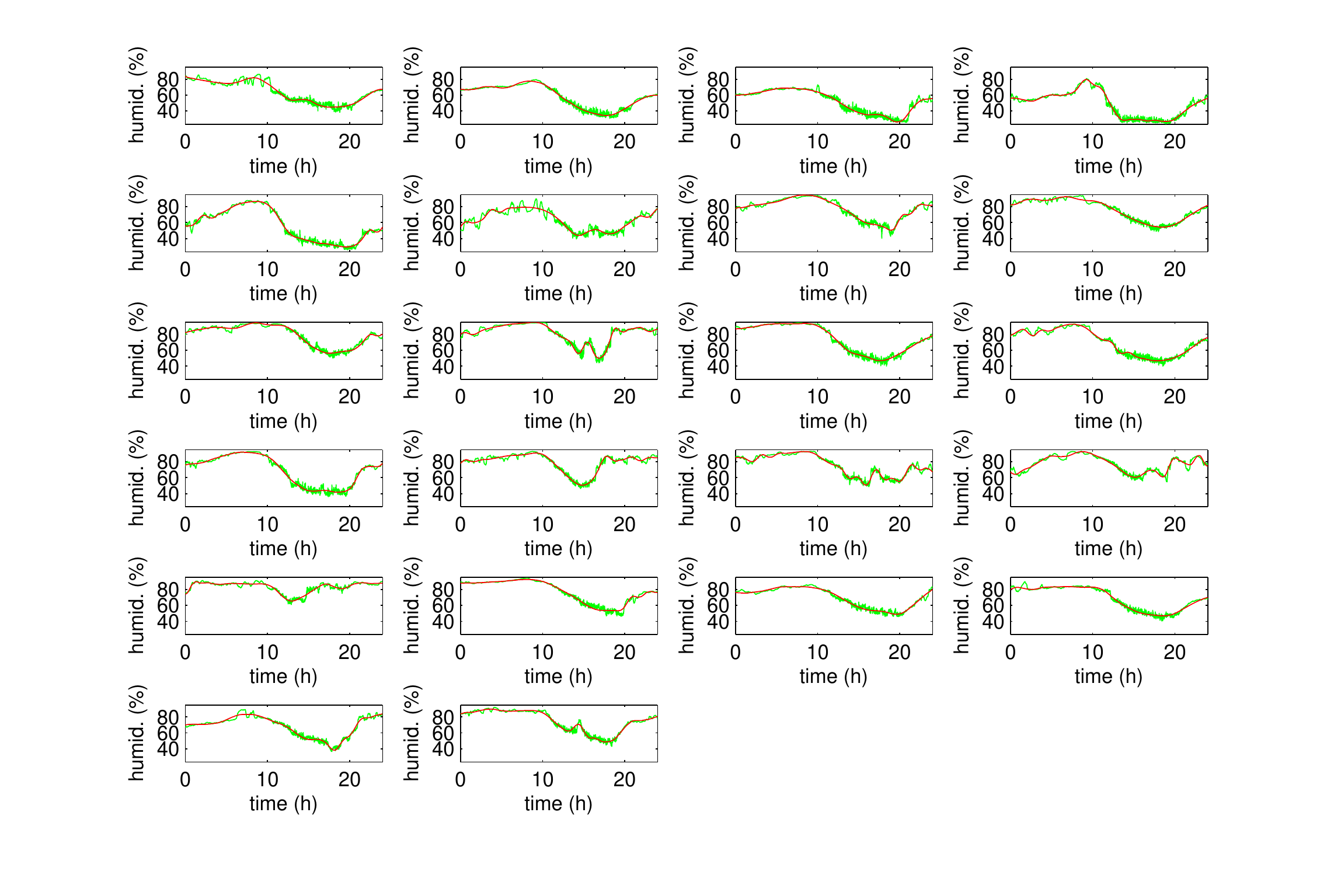}
\vspace*{-1.1cm} \caption{Bras\'{\i}lia's estimated and observed humidity curves for the month of March
2009.}\label{figumidmarBrasOS}
\end{figure}

\begin{figure}[!htb]
\centering
\includegraphics[width=16cm, height=8cm,height=8cm]{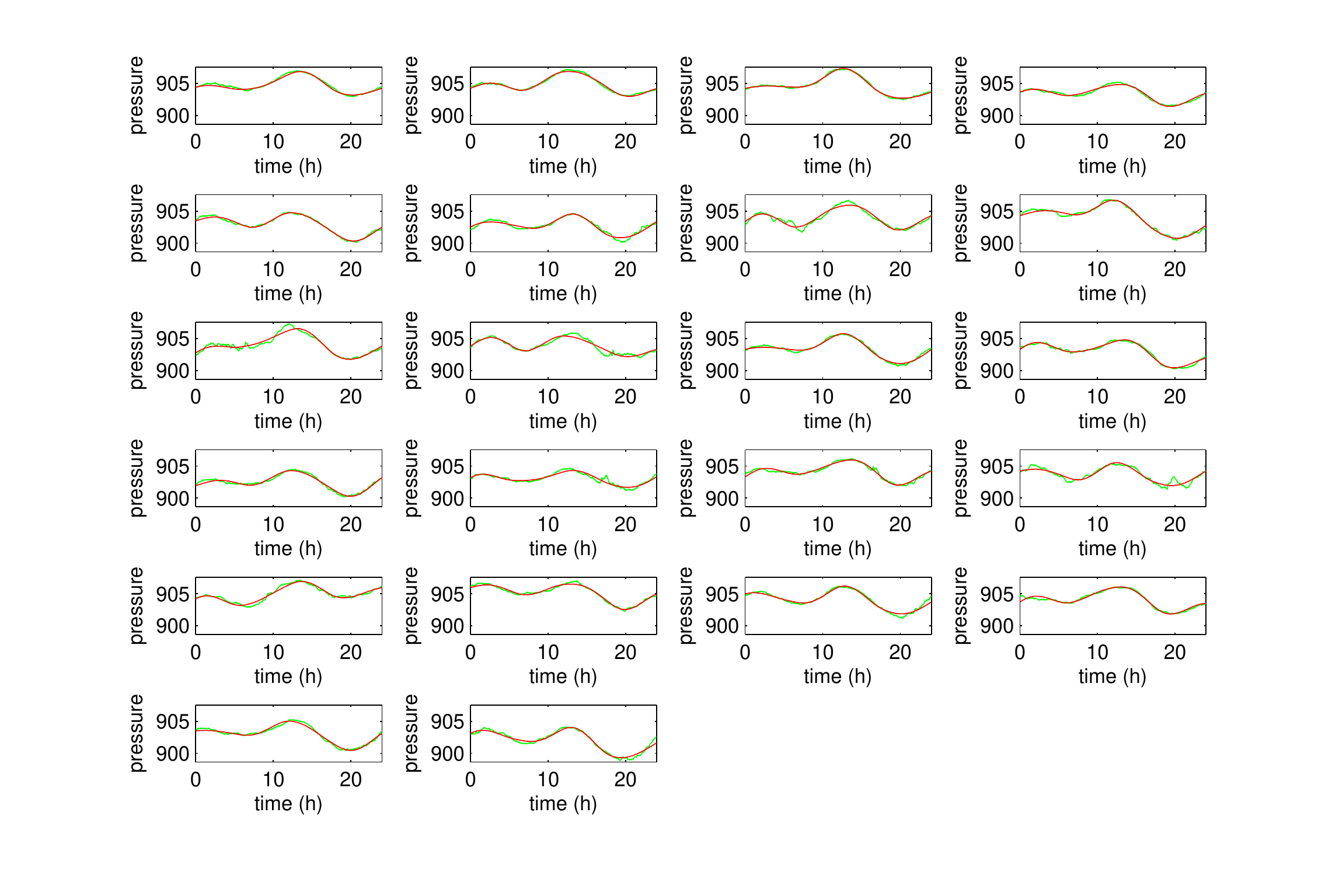}
\vspace*{-1.1cm} \caption{Bras\'{\i}lia's estimated and observed air pressure curves for the month of March
2009.}\label{figpresmarBrasOS}
\end{figure}

\begin{figure}[!htb]
\centering
\includegraphics[width=16cm, height=8cm, height=8cm]{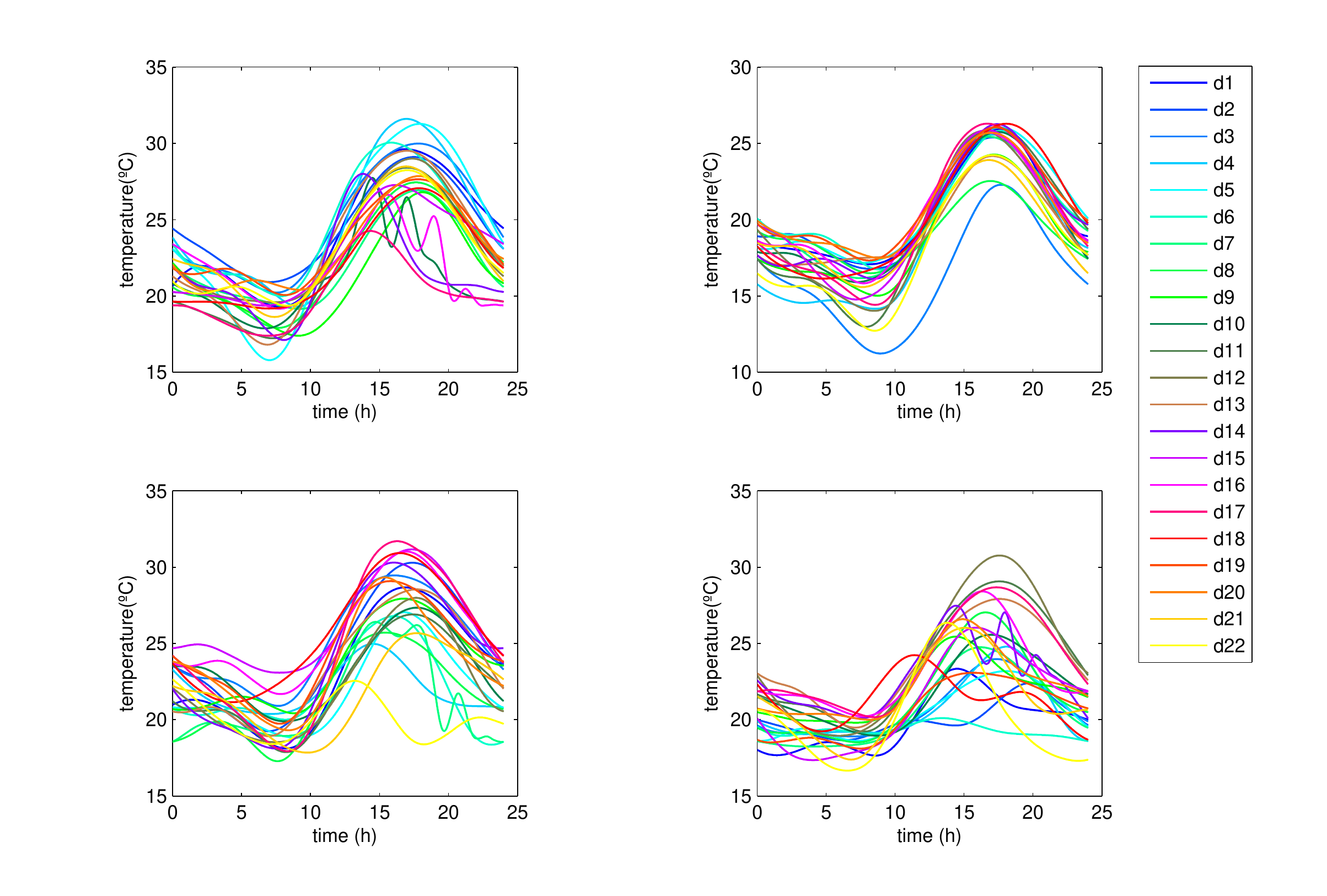}
\vspace*{-.6cm} \caption{Bras\'{\i}lia's Daily Estimated temperature curves for March, June, September and December of
2009. $d_i$ represents days $i$.}\label{figtempBras}
\end{figure}

\begin{figure}[!ht]
\centering
\includegraphics[width=16cm, height=8cm,height=8cm]{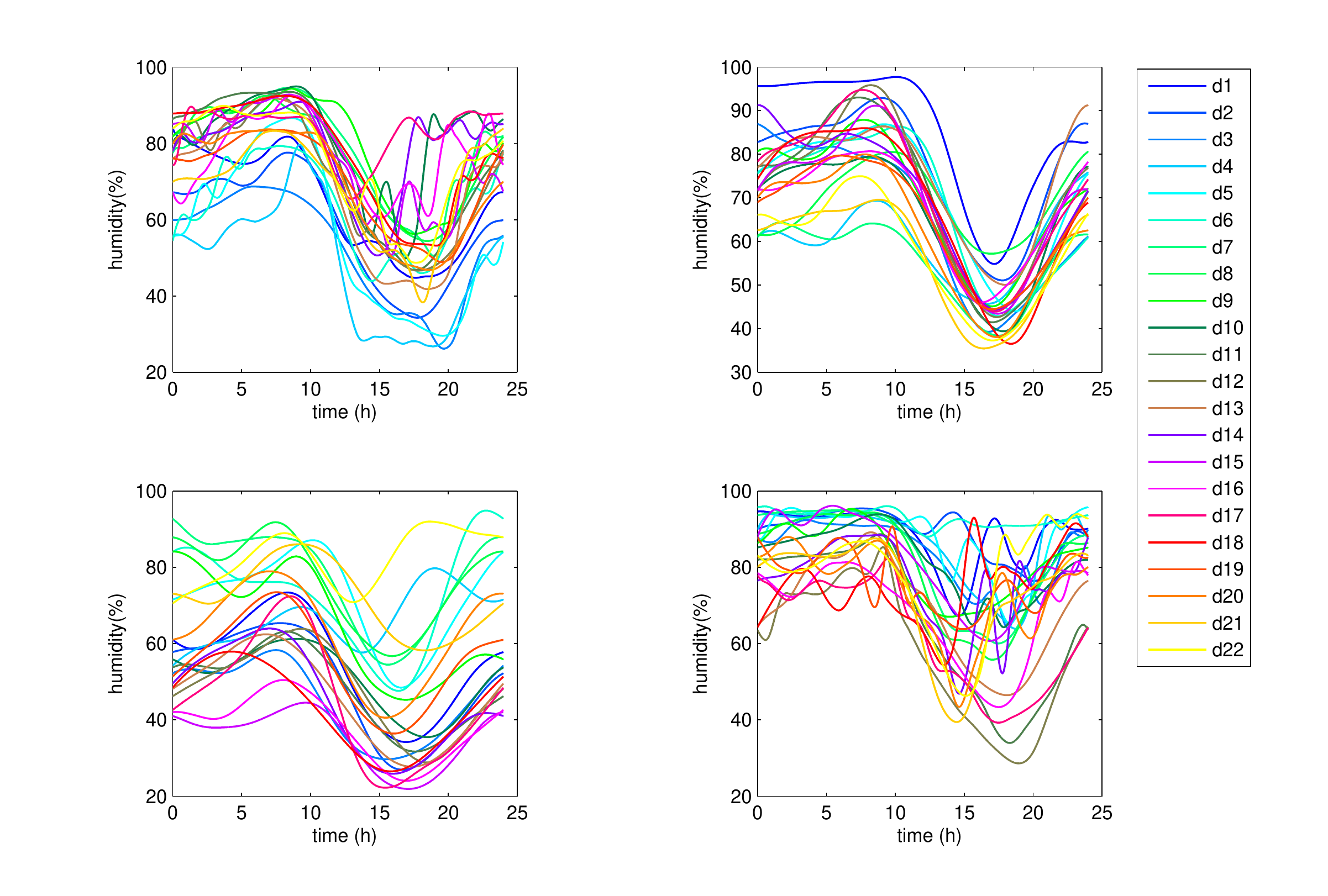}
\vspace*{-.6cm}  \caption{Bras\'{\i}lia's Daily Estimated humidity curves for March, June, September and December of
2009. $d_i$ represents days $i$.}\label{figumidBras}
\end{figure}

\begin{figure}[!hb]
\centering
\includegraphics[width=16cm, height=8cm,height=8cm]{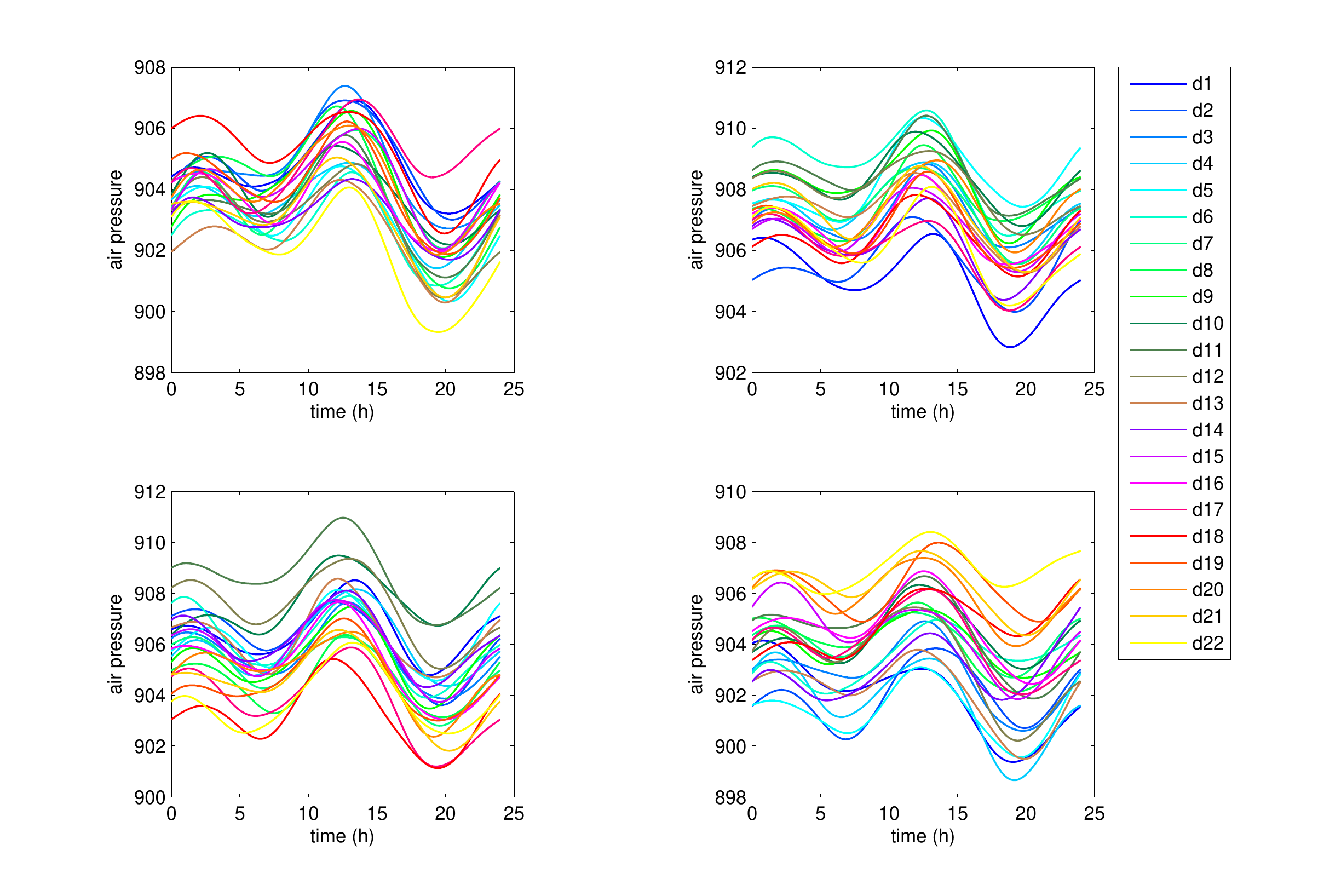}
\vspace*{-.6cm}  \caption{Bras\'{\i}lia's Daily Estimated atmospheric pressure curves for March, June, September and
December of 2009. $d_i$ represents days $i$.}\label{figpresBras}
\end{figure}

In the remaining text we analyze the data of Bras\'{\i}lia.  Figures \ref{figtempmarBrasOS}, \ref{figumidmarBrasOS} and \ref{figpresmarBrasOS} show the
temperature, humidity and atmospheric pressure observed and estimated the first 22 days of the month of March 2009 for
the city of Bras\'{\i}lia, respectively. A noisy curve of interpolated observed values and a smoothed estimated curve is presented for each day. 
Figures  \ref{figtempBras}, \ref{figumidBras} and \ref{figpresBras} present the estimates for the Bras\'{\i}lia's daily
curves of temperature, humidity and pressure, respectively. One sees some similar behavior on the daily temperature and
atmospheric pressure curves for each variable within each month, with some exceptions. In general, these daily curves
are very regular with one local minimum and maximum per day for the temperature and two local maxima and minima for the
pressure. The humidity curves present much wider daily amplitude and much less regular behavior within each month,
specially for the rainy season months, i.e. March and December.

\begin{figure}[!htb]
\centering
\includegraphics[width=16cm, height=8cm,height=8cm]{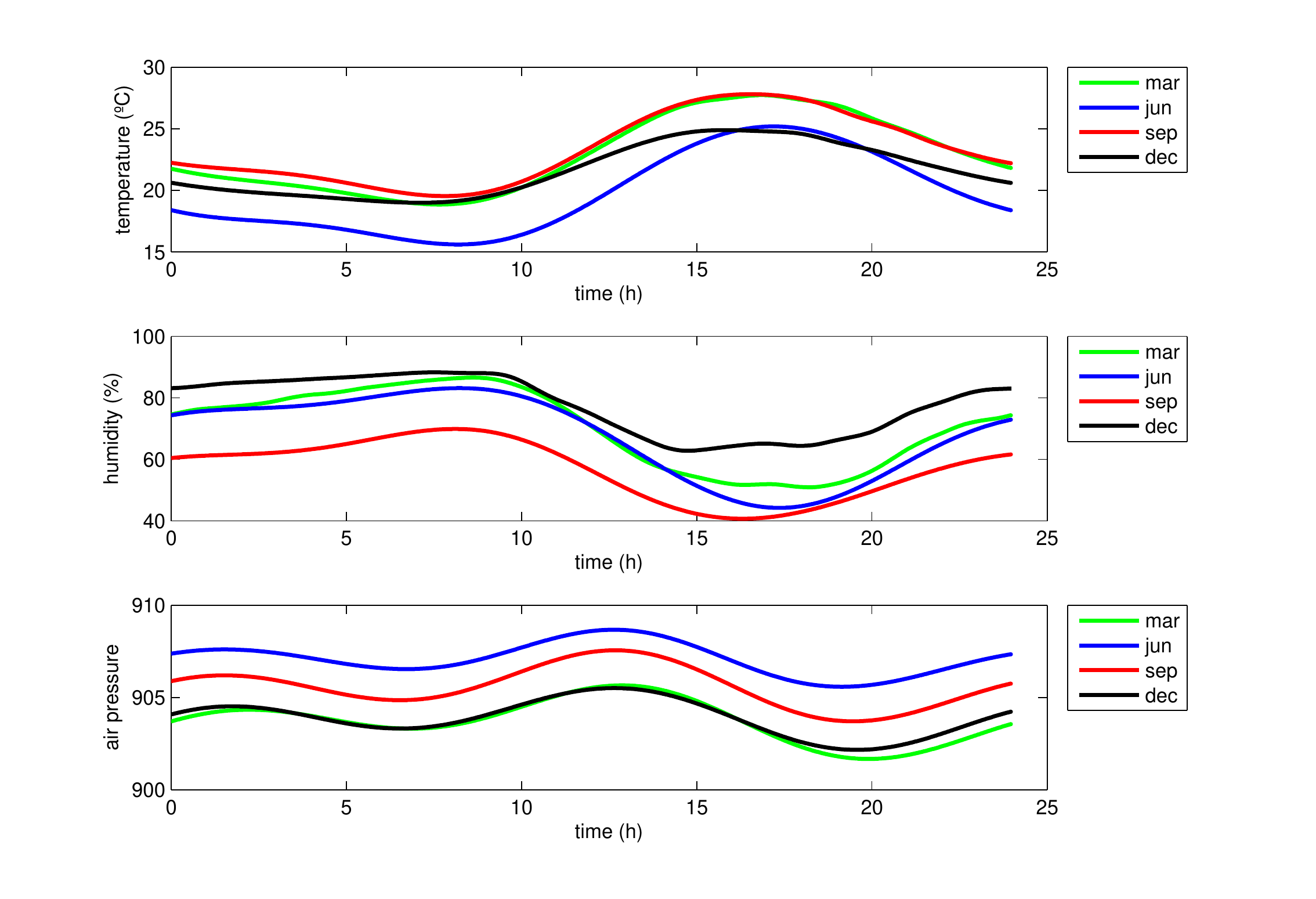}
\vspace*{-.6cm}  \caption{Bras\'{\i}lia's Mean Estimated curves for the first 22 days of March, June, September and
December of 2009. Separate Plots for mean temperature, humidity and pressure curves are presented.}\label{figmedias}
\end{figure}

Average Bras\'{\i}lia daily estimates curves for the first 22 days of March, June, September and December 2009 are shown in Figure \ref{figmedias}. One sees that June presents the coldest days, and September, the hottest days. The month of September is usually the driest, and December the wettest. Finally the months of March and December present lower atmospheric pressures compared to June and September.

To verify that the curves of climatic variables have identical behavior from one year to another applied the test
described in the proposed model to the observed data in 2009 and 2010 in Bras\'{\i}lia. Has taken  the observed curves
for June and September 2010 from Bras\'{\i}lia to perform this test. The data for March are incomplete and of December
were lost. Since the data are correlated, one can not perform the test developed in \citep{Abramovich2004} directly.
Therefore, before performing the test made the transformation $y_t-\hat{\rho}y_{t-1},\,\,t=1,2,\ldots,n$ in the
observed data for the uncorrelated errors. Thus, $y_t-\hat{\rho}y_{t-1}=f_t-\hat{\rho}f_{t-1}+u_t \Longleftrightarrow
z_t=g_t+u_t$, in which errors $u_t$ have approximately distribution  $\mathrm{N}(0,\sigma_u^2)$. To perform the test
replaces the function $f$ by the curve estimated for 2009. Thus we tested
\begin{equation*}
\text{H}_0:z-g\equiv\text{Constant} \quad \text{versus} \quad \text{H}_1:\left(z-g-\text{
Constant}\right)\in\mathcal{F}(\varrho),
\end{equation*}
with significance $\alpha = 5\%$. The test result for June and September to the station of Bras\'{\i}lia are in Table
\ref{tabTesteBras}.

\begin{table}[h]
  \centering
  \caption
  {Test results $\text{H}_0:z-g\equiv\text{Constant }  \text{ versus }
  \text{H}_1:\left(z-g-\text{Constant}\right)\in\mathcal{F}(\varrho)$
  with significance $\alpha = 5\%$.
  $T(j(6))+Q(j(6))$ is the value of statistics and $\sqrt{v_0^2(6)+w_0^2(6)}z_{0.95}$
  is the critical value}\label{tabTesteBras}

\begin{tabular}{ccccccc}
  \hline

 &  & Temperature & Humidity &   Pressure  \\
  \hline
\multirow{2}{*}{June}
& $T(j(6))+Q(j(6))$  & 213.43 & 81,900,00 & 215.69   \\
& $\sqrt{v_0^2(6)+w_0^2(6)}z_{0.95}$  & 0.22   & 2.99     & 0.08     \\
\hline \multirow{2}{*}{September}
& $T(j(6))+Q(j(6))$  & 466.01 & 3,396.40  & 43.94    \\
&  $\sqrt{v_0^2(6)+w_0^2(6)}z_{0.95}$  & 0.20   & 1.97     & 0.07     \\
  \hline
\end{tabular}

\end{table}

In all cases, the null hypothesis was rejected. Thus the curves observed in the months of June and September 2010 are
different curves observed in the same period of 2009.

\FloatBarrier

\section{Discussion}
The analysis of functional data has become more common in the last decades due to the exponential increase in computing power, which has driven the also increasing large datasets' acquisition and the development of appropriate statistical analysis tools. We present a modification of optimal wavelet procedures\citep{Abramovich2004} to deal with dependent errors. The theoretical advantages of correctly estimating the error dependence are shown, by simulation and application to real data set, to be also quite relevant in practice. 
   
\bibliographystyle{apalike}
\bibliography{Kistbib}

\newpage

\section*{Appendix}

We show below some figures and tables for the Doppler function simulation results. Results for the other functions were qualitatively equivalent, and are available as supplementary material. 
\begin{figure}[!htb]
\centering
\includegraphics[width=16cm, height=8cm]{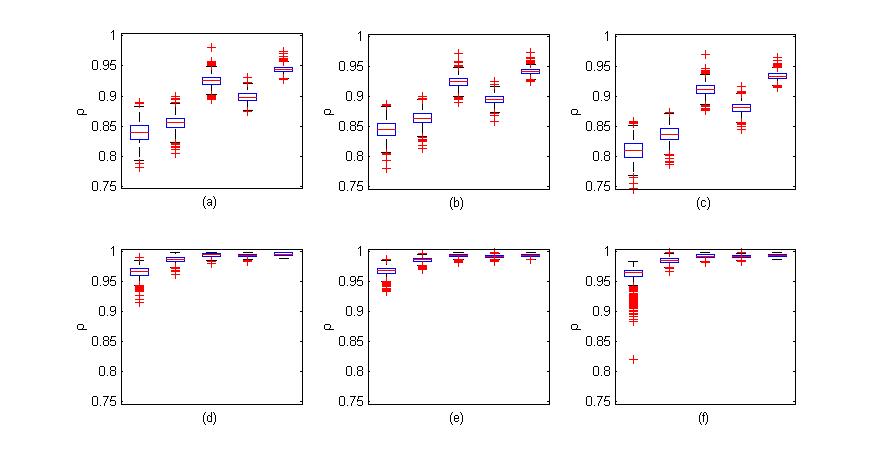}
\caption{Box-Plots for $\hat{\rho}$. Doppler function, $\rho=0.99$, $\mathrm{SNR}=1$. 1000 replications. 50 initial values randomly chosen from $U(-1,1)$.  Panels: (a) Linear functional step and 'db3';  (b)  Linear functional step and 'db6'; (c)   Linear functional step and 'sym8'; (d) Nonlinear functional step and 'db3';  (e)  Nonlinear functional step and 'db6'; (f)   Nonlinear functional step and 'sym8'.  In each panel box-plots for sample sizes $n=512;1024;2048;4096;8192$ are shown from left to right.}
\label{}
\end{figure}
\begin{figure}[!htb]
\centering
\includegraphics[width=16cm, height=8cm]{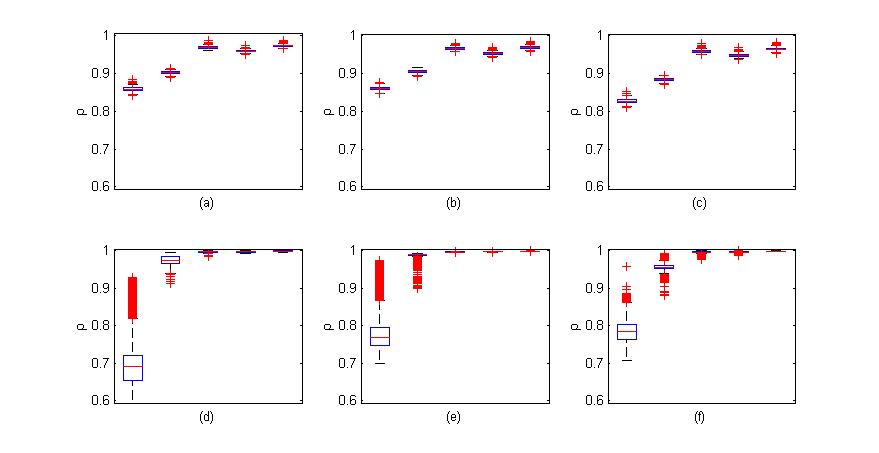}
 \caption{Box-Plots for $\hat{\rho}$. Doppler function, $\rho=0.99$, $\mathrm{SNR}=3$. 1000 replications. 50 initial values randomly chosen from $U(-1,1)$.  Panels: (a) Linear functional step and 'db3';  (b)  Linear functional step and 'db6'; (c)   Linear functional step and 'sym8'; (d) Nonlinear functional step and 'db3';  (e)  Nonlinear functional step and 'db6'; (f)   Nonlinear functional step and 'sym8'.  In each panel box-plots for sample sizes $n=512;1024;2048;4096;8192$ are shown from left to right.}
\label{}
\end{figure}
\begin{figure}[!htb]
\centering
\includegraphics[width=16cm, height=8cm]{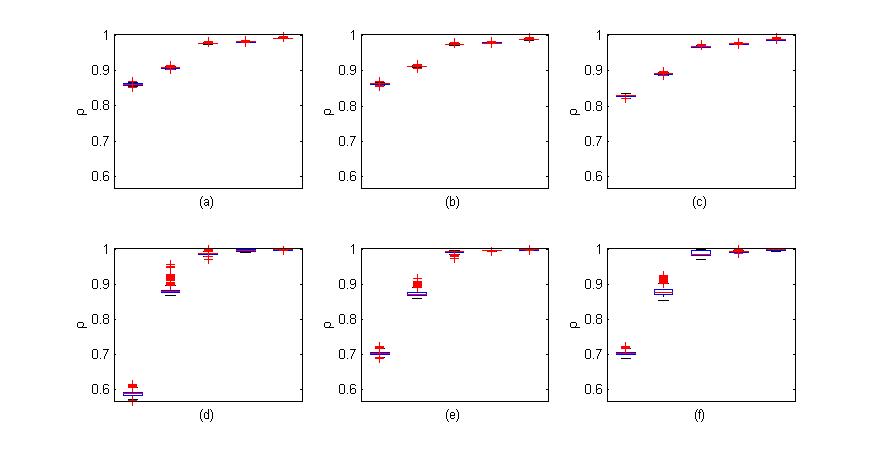}
 \caption{Box-Plots for $\hat{\rho}$. Doppler function, $\rho=0.99$, $\mathrm{SNR}=7$. 1000 replications. 50 initial values randomly chosen from $U(-1,1)$.  Panels: (a) Linear functional step and 'db3';  (b)  Linear functional step and 'db6'; (c)   Linear functional step and 'sym8'; (d) Nonlinear functional step and 'db3';  (e)  Nonlinear functional step and 'db6'; (f)   Nonlinear functional step and 'sym8'.  In each panel box-plots for sample sizes $n=512;1024;2048;4096;8192$ are shown from left to right.}
\label{}
\end{figure}
\begin{figure}[!htb]
\centering
\includegraphics[width=16cm, height=8cm]{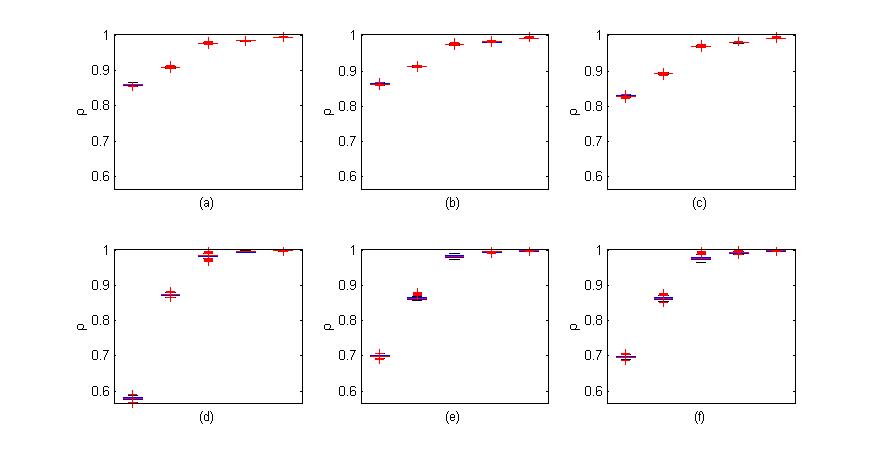}
\caption{Box-Plots for $\hat{\rho}$. Doppler function, $\rho=0.9999$, $\mathrm{SNR}=1$. 1000 replications. 50 initial values randomly chosen from $U(-1,1)$.  Panels: (a) Linear functional step and 'db3';  (b)  Linear functional step and 'db6'; (c)   Linear functional step and 'sym8'; (d) Nonlinear functional step and 'db3';  (e)  Nonlinear functional step and 'db6'; (f)   Nonlinear functional step and 'sym8'.  In each panel box-plots for sample sizes $n=512;1024;2048;4096;8192$ are shown from left to right.}
\label{}\end{figure}
\begin{figure}[!htb]
\centering
\includegraphics[width=16cm, height=8cm]{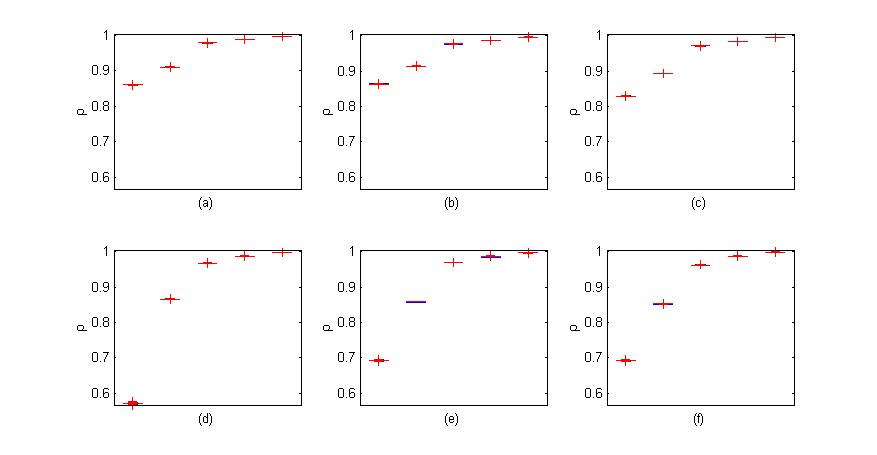}
\caption{Box-Plots for $\hat{\rho}$. Doppler function, $\rho=0.9999$, $\mathrm{SNR}=3$. 1000 replications. 50 initial values randomly chosen from $U(-1,1)$.  Panels: (a) Linear functional step and 'db3';  (b)  Linear functional step and 'db6'; (c)   Linear functional step and 'sym8'; (d) Nonlinear functional step and 'db3';  (e)  Nonlinear functional step and 'db6'; (f)   Nonlinear functional step and 'sym8'.  In each panel box-plots for sample sizes $n=512;1024;2048;4096;8192$ are shown from left to right.}
\label{}\end{figure}
\begin{figure}[!htb]
\centering
\includegraphics[width=16cm, height=8cm]{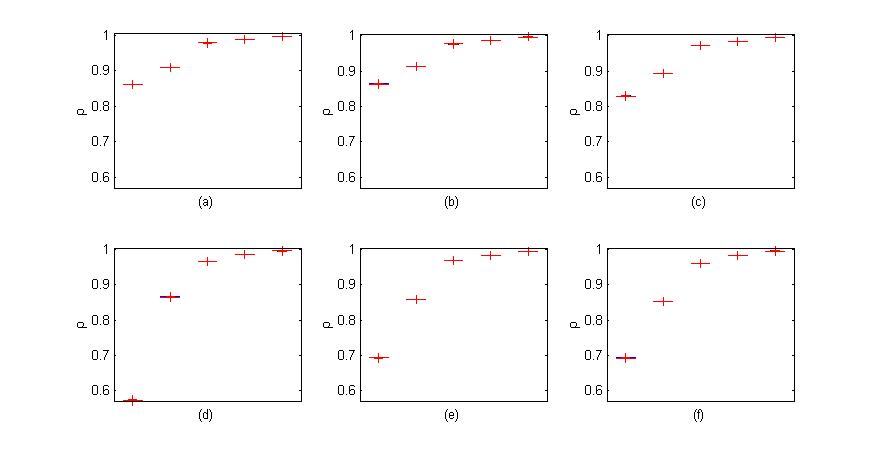}
\caption{Box-Plots for $\hat{\rho}$. Doppler function, $\rho=0.9999$, $\mathrm{SNR}=7$. 1000 replications. 50 initial values randomly chosen from $U(-1,1)$.  Panels: (a) Linear functional step and 'db3';  (b)  Linear functional step and 'db6'; (c)   Linear functional step and 'sym8'; (d) Nonlinear functional step and 'db3';  (e)  Nonlinear functional step and 'db6'; (f)   Nonlinear functional step and 'sym8'.  In each panel box-plots for sample sizes $n=512;1024;2048;4096;8192$ are shown from left to right.}
\label{}
\end{figure}
\begin{figure}[!htb]
\centering
\includegraphics[width=16cm, height=8cm]{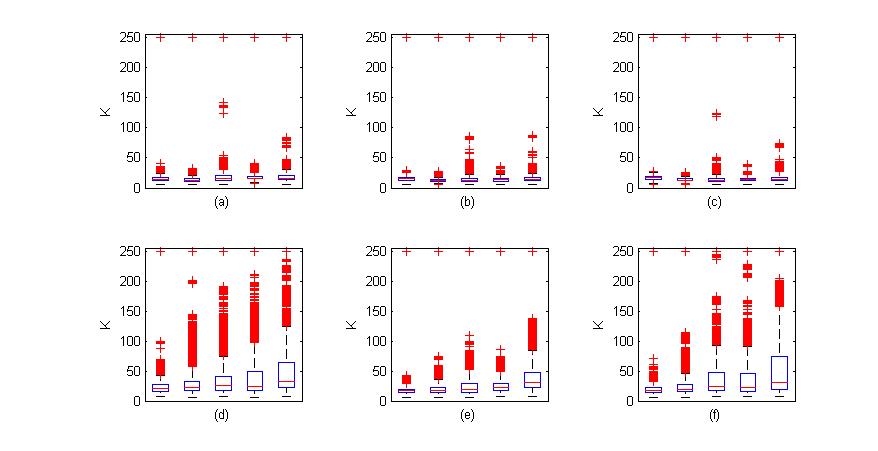}
 \caption{Box-Plots for the number of iterations until numerical convergence. Doppler function, $\rho=0.99$, $\mathrm{SNR}=1$. 1000 replications. 50 initial values randomly chosen from $U(-1,1)$.  Panels: (a) Linear functional step and 'db3';  (b)  Linear functional step and 'db6'; (c)   Linear functional step and 'sym8'; (d) Nonlinear functional step and 'db3';  (e)  Nonlinear functional step and 'db6'; (f)   Nonlinear functional step and 'sym8'.  In each panel box-plots for sample sizes $n=512;1024;2048;4096;8192$ are shown from left to right.}
\label{}\end{figure}
\begin{figure}[!htb]
\centering
\includegraphics[width=16cm, height=8cm]{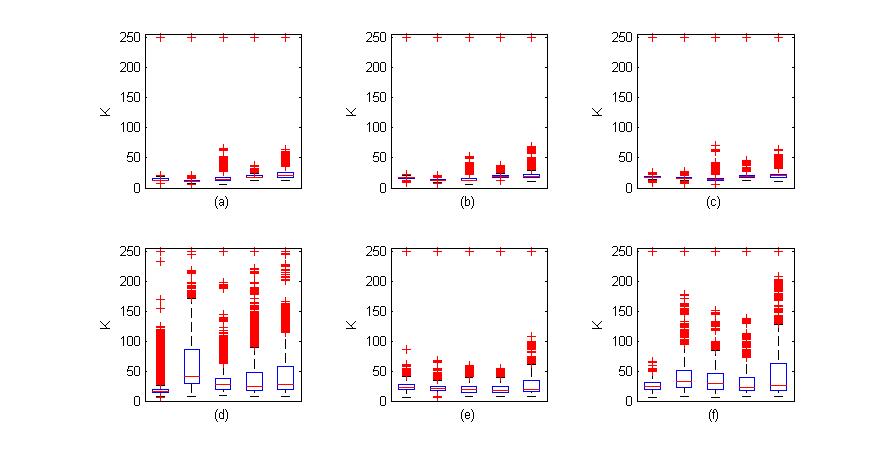}
 \caption{Box-Plots  for the number of iterations until numerical convergence. Doppler function, $\rho=0.99$, $\mathrm{SNR}=3$. 1000 replications. 50 initial values randomly chosen from $U(-1,1)$.  Panels: (a) Linear functional step and 'db3';  (b)  Linear functional step and 'db6'; (c)   Linear functional step and 'sym8'; (d) Nonlinear functional step and 'db3';  (e)  Nonlinear functional step and 'db6'; (f)   Nonlinear functional step and 'sym8'.  In each panel box-plots for sample sizes $n=512;1024;2048;4096;8192$ are shown from left to right.}
\label{}\end{figure}
\begin{figure}[!htb]
\centering
\includegraphics[width=16cm, height=8cm]{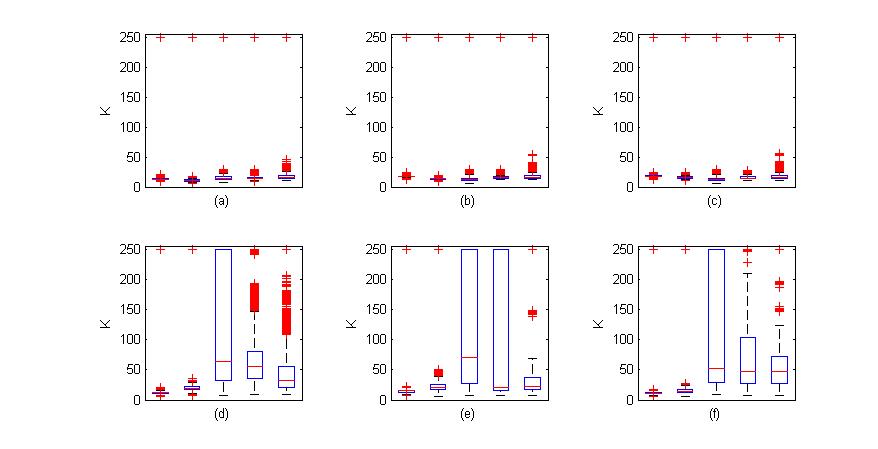}
 \caption{Box-Plots for  for the number of iterations until numerical convergence. Doppler function, $\rho=0.99$, $\mathrm{SNR}=7$. 1000 replications. 50 initial values randomly chosen from $U(-1,1)$.  Panels: (a) Linear functional step and 'db3';  (b)  Linear functional step and 'db6'; (c)   Linear functional step and 'sym8'; (d) Nonlinear functional step and 'db3';  (e)  Nonlinear functional step and 'db6'; (f)   Nonlinear functional step and 'sym8'.  In each panel box-plots for sample sizes $n=512;1024;2048;4096;8192$ are shown from left to right.}
\label{}
\end{figure}
\begin{figure}[!htb]
\centering
\includegraphics[width=16cm, height=8cm]{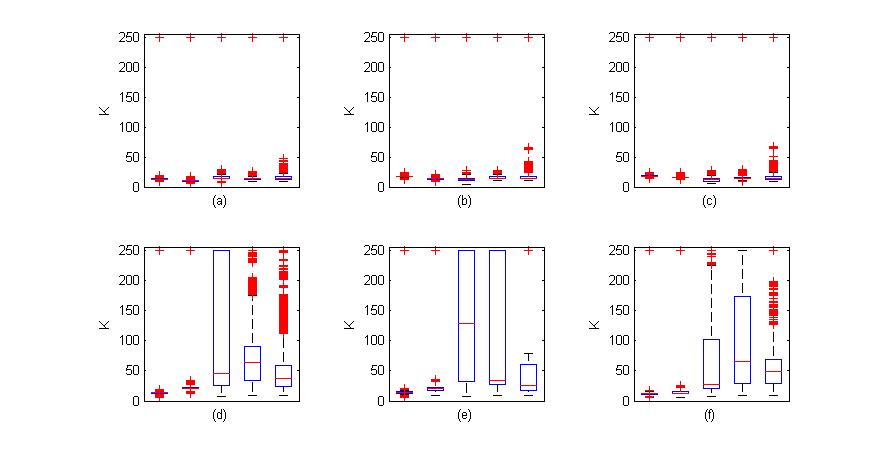}
 \caption{Box-Plots for the number of iterations until numerical convergence. Doppler function, $\rho=0.9999$, $\mathrm{SNR}=1$. 1000 replications. 50 initial values randomly chosen from $U(-1,1)$.  Panels: (a) Linear functional step and 'db3';  (b)  Linear functional step and 'db6'; (c)   Linear functional step and 'sym8'; (d) Nonlinear functional step and 'db3';  (e)  Nonlinear functional step and 'db6'; (f)   Nonlinear functional step and 'sym8'.  In each panel box-plots for sample sizes $n=512;1024;2048;4096;8192$ are shown from left to right.}
\label{}\end{figure}
\begin{figure}[!htb]
\centering
\includegraphics[width=16cm, height=8cm]{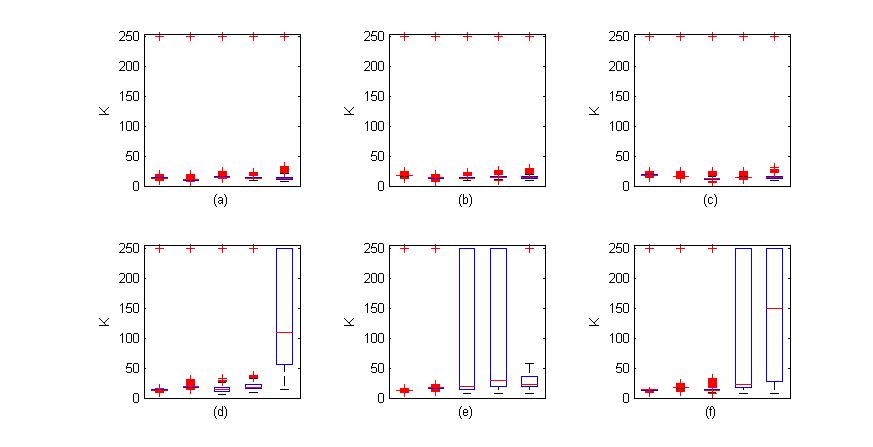}
 \caption{Box-Plots  for the number of iterations until numerical convergence. Doppler function, $\rho=0.9999$, $\mathrm{SNR}=3$. 1000 replications. 50 initial values randomly chosen from $U(-1,1)$.  Panels: (a) Linear functional step and 'db3';  (b)  Linear functional step and 'db6'; (c)   Linear functional step and 'sym8'; (d) Nonlinear functional step and 'db3';  (e)  Nonlinear functional step and 'db6'; (f)   Nonlinear functional step and 'sym8'.  In each panel box-plots for sample sizes $n=512;1024;2048;4096;8192$ are shown from left to right.}
\label{}\end{figure}
\begin{figure}[!htb]
\centering
\includegraphics[width=16cm, height=8cm]{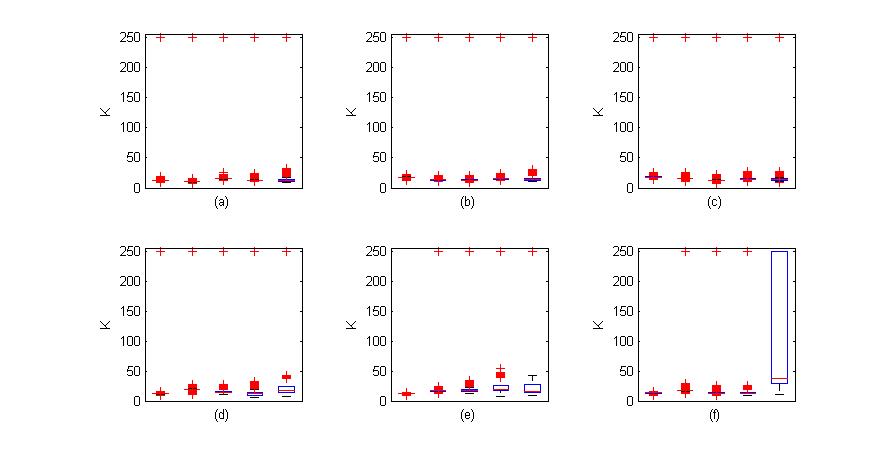}
 \caption{Box-Plots for  for the number of iterations until numerical convergence. Doppler function, $\rho=0.9999$, $\mathrm{SNR}=7$. 1000 replications. 50 initial values randomly chosen from $U(-1,1)$.  Panels: (a) Linear functional step and 'db3';  (b)  Linear functional step and 'db6'; (c)   Linear functional step and 'sym8'; (d) Nonlinear functional step and 'db3';  (e)  Nonlinear functional step and 'db6'; (f)   Nonlinear functional step and 'sym8'.  In each panel box-plots for sample sizes $n=512;1024;2048;4096;8192$ are shown from left to right.}
\label{}
\end{figure}
\begin{figure}[!htb]
\centering
\includegraphics[width=16cm, height=8cm]{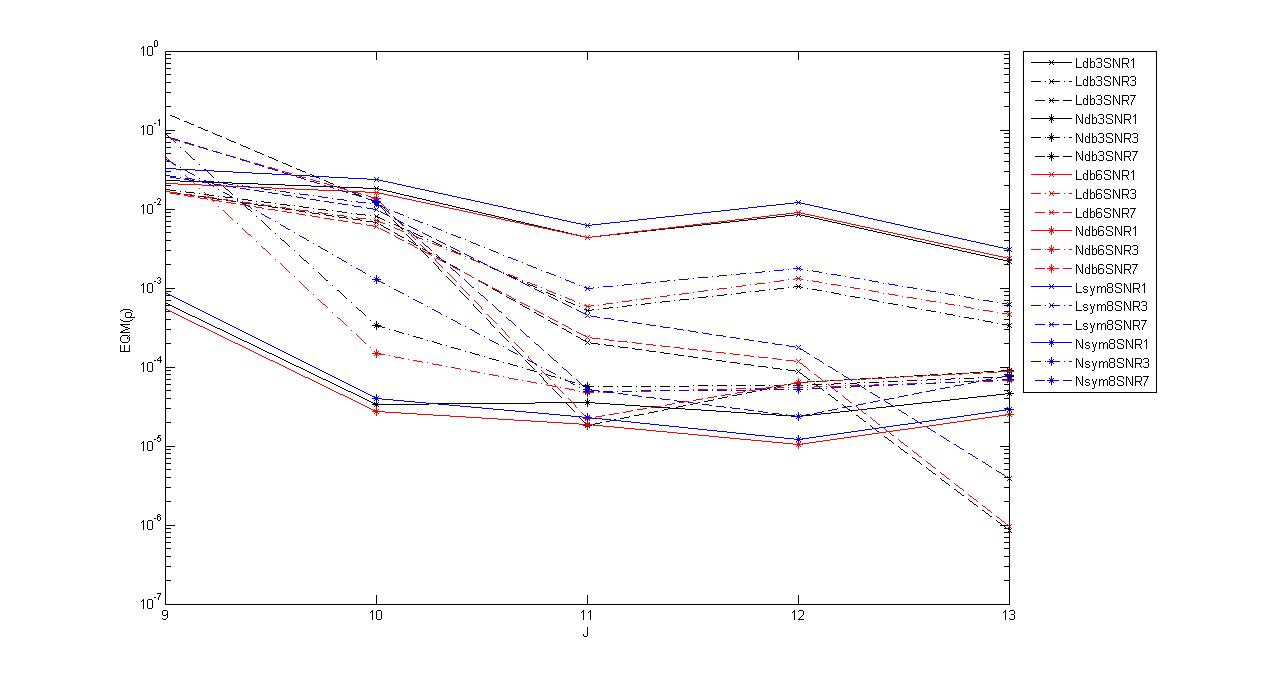}
\caption{$MSE(\hat{\rho})$ as a function of sample sizes $n=512;1024;2048;4096;8192$.  Doppler function, $\rho=0.99$. 1000 replications. 50 initial values randomly chosen from $U(-1,1)$.  18 plots are labeled by xbasisSNRy, where: x='L' or 'N' for linear and nonlinear steps respectively; basis='db3', 'db6' or 'sym8'; y=1,3 or 7. }
\label{}
\end{figure}'
\begin{figure}[!htb]
\centering
\includegraphics[width=16cm, height=8cm]{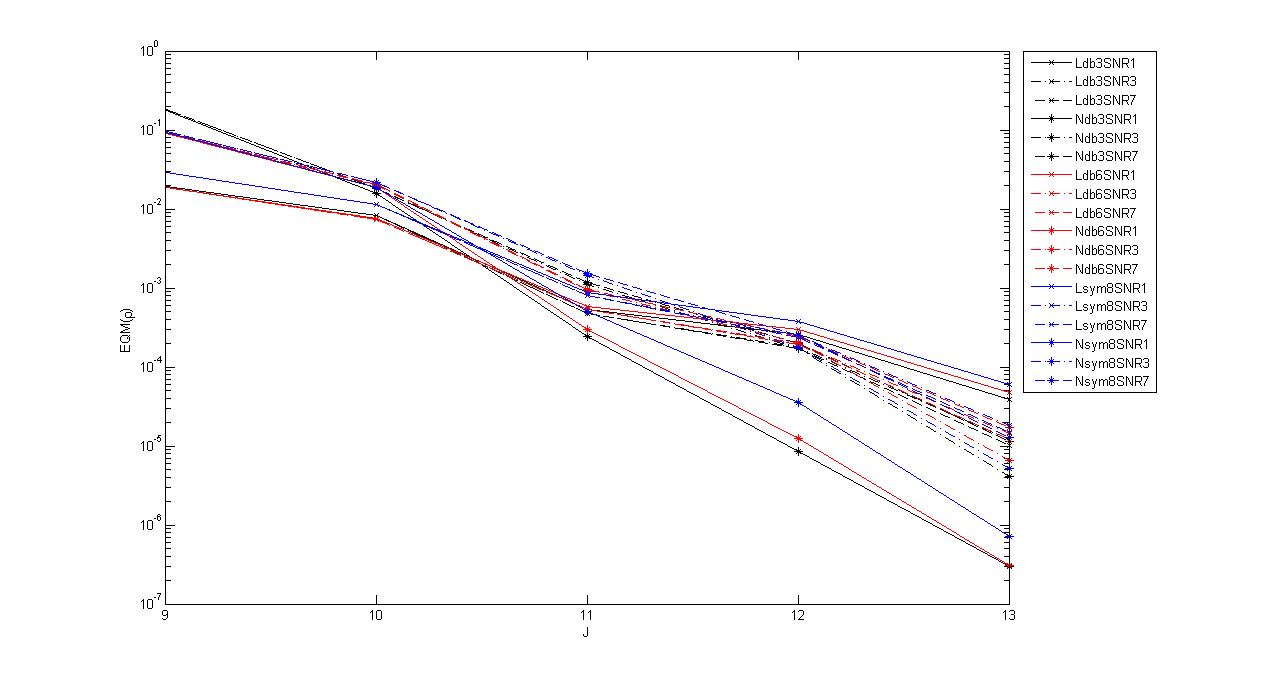}
\caption{$MSE(\hat{\rho})$ as a function of sample sizes $n=512;1024;2048;4096;8192$.  Doppler function, $\rho=0.9999$. 1000 replications. 50 initial values randomly chosen from $U(-1,1)$.  18 plots are labeled by xbasisSNRy, where: x='L' or 'N' for linear and nonlinear steps respectively; basis='db3', 'db6' or 'sym8'; y=1,3 or 7. }
\label{}
\end{figure}
\FloatBarrier
\begin{sidewaystable}
\centering
 { \footnotesize
\caption{Integrated Squared Mean Error for $\hat{f}$. Doppler function, $\rho=0.99$, $\mathrm{SNR}=1,3,7$. 1000 replications and sample sizes $n=512;1024;2048;4096;8192$. 50 initial values randomly chosen from $U(-1,1)$.  Wavelet bases: 'db3', db6' and 'sym8'.  Functional estimation steps are called 'L", 'NL' or 'NB' if linear, nonlinear with term by term thresholding or nonlinear with block thresholding is employed, respectively. $IMSE(xA)$ represent the average IMSE of $\hat{f}$ for the 1000 replication, 50 initial values in the combination of an SNR=x and thresholding procedure A, where  x=1,3 or 7 and A= 'L', 'NL' or 'NB'.}
\label{rhoresultseqmf1_9900}
\input{Mfresultseqmf1_9900.tex} }
\end{sidewaystable}
\begin{sidewaystable}
\centering
 { \footnotesize
\caption{Integrated Squared Mean Error for $\hat{f}$. Doppler function, $\rho=0.9999$, $\mathrm{SNR}=1,3,7$. 1000 replications and sample sizes $n=512;1024;2048;4096;8192$. 50 initial values randomly chosen from $U(-1,1)$.  Wavelet bases: 'db3', db6' and 'sym8'.  Functional estimation steps are called 'L", 'NL' or 'NB' if linear, nonlinear with term by term thresholding or nonlinear with block thresholding is employed, respectively. $IMSE(xA)$ represent the average IMSE of $\hat{f}$ for the 1000 replication, 50 initial values in the combination of an SNR=x and thresholding procedure A, where  x=1,3 or 7 and A= 'L', 'NL' or 'NB'.}\label{rhoresultseqmf1_9999}
\input{Mfresultseqmf1_9999.tex}
}
\end{sidewaystable}
\end{document}

%% file: Mfresultseqmf1_9900.tex
\begin{tabular}{cc|rrrrr|rrrrr|rrrrr}
\hline
& & \multicolumn{5}{|c|}{SNR=1} & \multicolumn{5}{|c|}{SNR=3} & \multicolumn{5}{|c}{SNR=7} \\
\hline
 & $n$ & 512 & \hspace{-.2cm}1024 & \hspace{-.2cm}2048 & \hspace{-.2cm}4096 & \hspace{-.2cm}8192 & 512 & \hspace{-.2cm}1024 & \hspace{-.2cm}2048 & \hspace{-.2cm}4096 & \hspace{-.2cm}8192 & 512 & \hspace{-.2cm}1024 & \hspace{-.2cm}2048 & \hspace{-.2cm}4096 & \hspace{-.2cm}8192    \\
\hline
 \multirow{9}{*}{db3 \hspace{-.3cm}}
 & $IMSE(1L)$& \hspace{-.2cm} 0.104   & \hspace{-.2cm} 0.052  & \hspace{-.2cm} 0.026  & \hspace{-.2cm} 0.012  & \hspace{-.2cm} 0.006  & \hspace{-.2cm} 0.155  & \hspace{-.2cm} 0.078  & \hspace{-.2cm} 0.040  & \hspace{-.2cm} 0.015  & \hspace{-.2cm} 0.007  & \hspace{-.2cm} 0.408  & \hspace{-.2cm} 0.205  & \hspace{-.2cm} 0.111  & \hspace{-.2cm} 0.028  & \hspace{-.2cm} 0.014  \\
 & $IMSE(2L)$& \hspace{-.2cm} 0.104   & \hspace{-.2cm} 0.052  & \hspace{-.2cm} 0.026  & \hspace{-.2cm} 0.012  & \hspace{-.2cm} 0.006  & \hspace{-.2cm} 0.155  & \hspace{-.2cm} 0.078  & \hspace{-.2cm} 0.040  & \hspace{-.2cm} 0.015  & \hspace{-.2cm} 0.007  & \hspace{-.2cm} 0.408  & \hspace{-.2cm} 0.205  & \hspace{-.2cm} 0.111  & \hspace{-.2cm} 0.028  & \hspace{-.2cm} 0.014  \\
 & $IMSE(3L)$& \hspace{-.2cm} 0.104   & \hspace{-.2cm} 0.052  & \hspace{-.2cm} 0.026  & \hspace{-.2cm} 0.012  & \hspace{-.2cm} 0.006  & \hspace{-.2cm} 0.155  & \hspace{-.2cm} 0.078  & \hspace{-.2cm} 0.040  & \hspace{-.2cm} 0.015  & \hspace{-.2cm} 0.007  & \hspace{-.2cm} 0.408  & \hspace{-.2cm} 0.205  & \hspace{-.2cm} 0.111  & \hspace{-.2cm} 0.028  & \hspace{-.2cm} 0.014  \\
 & $IMSE(1NT)$& \hspace{-.2cm} 0.151  & \hspace{-.2cm} 0.138  & \hspace{-.2cm} 0.372  & \hspace{-.2cm} 0.099  & \hspace{-.2cm} 0.225  & \hspace{-.2cm} 0.108  & \hspace{-.2cm} 0.170  & \hspace{-.2cm} 0.748  & \hspace{-.2cm} 0.236  & \hspace{-.2cm} 0.432  & \hspace{-.2cm} 0.118  & \hspace{-.2cm} 0.061  & \hspace{-.2cm} 0.383  & \hspace{-.2cm} 0.540  & \hspace{-.2cm} 1.137  \\
 & $IMSE(2NT)$& \hspace{-.2cm} 0.151  & \hspace{-.2cm} 0.138  & \hspace{-.2cm} 0.372  & \hspace{-.2cm} 0.099  & \hspace{-.2cm} 0.225  & \hspace{-.2cm} 0.108  & \hspace{-.2cm} 0.170  & \hspace{-.2cm} 0.748  & \hspace{-.2cm} 0.236  & \hspace{-.2cm} 0.432  & \hspace{-.2cm} 0.118  & \hspace{-.2cm} 0.061  & \hspace{-.2cm} 0.382  & \hspace{-.2cm} 0.540  & \hspace{-.2cm} 1.137  \\
 & $IMSE(3NT)$& \hspace{-.2cm} 0.151  & \hspace{-.2cm} 0.138  & \hspace{-.2cm} 0.372  & \hspace{-.2cm} 0.099  & \hspace{-.2cm} 0.225  & \hspace{-.2cm} 0.109  & \hspace{-.2cm} 0.170  & \hspace{-.2cm} 0.748  & \hspace{-.2cm} 0.236  & \hspace{-.2cm} 0.432  & \hspace{-.2cm} 0.118  & \hspace{-.2cm} 0.061  & \hspace{-.2cm} 0.380  & \hspace{-.2cm} 0.540  & \hspace{-.2cm} 1.137  \\
 & $IMSE(1NB)$& \hspace{-.2cm} 0.107  & \hspace{-.2cm} 0.068  & \hspace{-.2cm} 0.089  & \hspace{-.2cm} 0.033  & \hspace{-.2cm} 0.054  & \hspace{-.2cm} 0.094  & \hspace{-.2cm} 0.050  & \hspace{-.2cm} 0.029  & \hspace{-.2cm} 0.023  & \hspace{-.2cm} 0.025  & \hspace{-.2cm} 0.093  & \hspace{-.2cm} 0.045  & \hspace{-.2cm} 0.027  & \hspace{-.2cm} 0.013  & \hspace{-.2cm} 0.008  \\
 & $IMSE(2NB)$& \hspace{-.2cm} 0.107  & \hspace{-.2cm} 0.068  & \hspace{-.2cm} 0.089  & \hspace{-.2cm} 0.033  & \hspace{-.2cm} 0.054  & \hspace{-.2cm} 0.094  & \hspace{-.2cm} 0.050  & \hspace{-.2cm} 0.029  & \hspace{-.2cm} 0.023  & \hspace{-.2cm} 0.025  & \hspace{-.2cm} 0.093  & \hspace{-.2cm} 0.045  & \hspace{-.2cm} 0.027  & \hspace{-.2cm} 0.013  & \hspace{-.2cm} \textbf{0.008}  \\
 & $IMSE(3NB)$& \hspace{-.2cm} 0.107  & \hspace{-.2cm} 0.068  & \hspace{-.2cm} 0.089  & \hspace{-.2cm} 0.033  & \hspace{-.2cm} 0.054  & \hspace{-.2cm} 0.094  & \hspace{-.2cm} 0.050  & \hspace{-.2cm} \textbf{0.029}  & \hspace{-.2cm} 0.023  & \hspace{-.2cm} 0.025  & \hspace{-.2cm} 0.093  & \hspace{-.2cm} 0.045  & \hspace{-.2cm} 0.027  & \hspace{-.2cm} 0.013  & \hspace{-.2cm} 0.008  \\
\hline
 \multirow{9}{*}{db6 \hspace{-.3cm}}
 & $IMSE(1L)$& \hspace{-.2cm} 0.103   & \hspace{-.2cm} 0.050  & \hspace{-.2cm} 0.024  & \hspace{-.2cm} \textbf{0.011}  & \hspace{-.2cm} \textbf{0.005}  & \hspace{-.2cm} 0.167  & \hspace{-.2cm} 0.078  & \hspace{-.2cm} 0.038  & \hspace{-.2cm} 0.014  & \hspace{-.2cm} 0.006  & \hspace{-.2cm} 0.483  & \hspace{-.2cm} 0.217  & \hspace{-.2cm} 0.107  & \hspace{-.2cm} 0.026  & \hspace{-.2cm} 0.013  \\
 & $IMSE(2L)$& \hspace{-.2cm} 0.103   & \hspace{-.2cm} 0.050  & \hspace{-.2cm} 0.024  & \hspace{-.2cm} 0.011  & \hspace{-.2cm} 0.005  & \hspace{-.2cm} 0.167  & \hspace{-.2cm} 0.078  & \hspace{-.2cm} 0.038  & \hspace{-.2cm} 0.014  & \hspace{-.2cm} 0.006  & \hspace{-.2cm} 0.483  & \hspace{-.2cm} 0.217  & \hspace{-.2cm} 0.107  & \hspace{-.2cm} 0.026  & \hspace{-.2cm} 0.013  \\
 & $IMSE(3L)$& \hspace{-.2cm} 0.103   & \hspace{-.2cm} 0.050  & \hspace{-.2cm} 0.024  & \hspace{-.2cm} 0.011  & \hspace{-.2cm} 0.005  & \hspace{-.2cm} 0.167  & \hspace{-.2cm} 0.078  & \hspace{-.2cm} 0.038  & \hspace{-.2cm} 0.014  & \hspace{-.2cm} 0.006  & \hspace{-.2cm} 0.483  & \hspace{-.2cm} 0.217  & \hspace{-.2cm} 0.107  & \hspace{-.2cm} 0.026  & \hspace{-.2cm} 0.013  \\
 & $IMSE(1NT)$& \hspace{-.2cm} 0.151  & \hspace{-.2cm} 0.078  & \hspace{-.2cm} 0.065  & \hspace{-.2cm} 0.017  & \hspace{-.2cm} 0.018  & \hspace{-.2cm} 0.119  & \hspace{-.2cm} 0.296  & \hspace{-.2cm} 0.248  & \hspace{-.2cm} 0.077  & \hspace{-.2cm} 0.051  & \hspace{-.2cm} 0.131  & \hspace{-.2cm} 0.058  & \hspace{-.2cm} 0.375  & \hspace{-.2cm} 0.191  & \hspace{-.2cm} 0.184  \\
 & $IMSE(2NT)$& \hspace{-.2cm} 0.151  & \hspace{-.2cm} 0.078  & \hspace{-.2cm} 0.065  & \hspace{-.2cm} 0.017  & \hspace{-.2cm} 0.018  & \hspace{-.2cm} 0.119  & \hspace{-.2cm} 0.296  & \hspace{-.2cm} 0.248  & \hspace{-.2cm} 0.077  & \hspace{-.2cm} 0.051  & \hspace{-.2cm} 0.131  & \hspace{-.2cm} 0.058  & \hspace{-.2cm} 0.380  & \hspace{-.2cm} 0.192  & \hspace{-.2cm} 0.184  \\
 & $IMSE(3NT)$& \hspace{-.2cm} 0.151  & \hspace{-.2cm} 0.078  & \hspace{-.2cm} 0.065  & \hspace{-.2cm} 0.017  & \hspace{-.2cm} 0.018  & \hspace{-.2cm} 0.121  & \hspace{-.2cm} 0.297  & \hspace{-.2cm} 0.248  & \hspace{-.2cm} 0.077  & \hspace{-.2cm} 0.051  & \hspace{-.2cm} 0.131  & \hspace{-.2cm} 0.058  & \hspace{-.2cm} 0.385  & \hspace{-.2cm} 0.193  & \hspace{-.2cm} 0.184  \\
 & $IMSE(1NB)$& \hspace{-.2cm} 0.103  & \hspace{-.2cm} 0.059  & \hspace{-.2cm} 0.048  & \hspace{-.2cm} 0.023  & \hspace{-.2cm} 0.021  & \hspace{-.2cm} 0.093  & \hspace{-.2cm} 0.050  & \hspace{-.2cm} 0.084  & \hspace{-.2cm} 0.082  & \hspace{-.2cm} 0.066  & \hspace{-.2cm} 0.093  & \hspace{-.2cm} 0.045  & \hspace{-.2cm} 0.024  & \hspace{-.2cm} 0.011  & \hspace{-.2cm} 0.072  \\
 & $IMSE(2NB)$& \hspace{-.2cm} 0.103  & \hspace{-.2cm} 0.059  & \hspace{-.2cm} 0.048  & \hspace{-.2cm} 0.023  & \hspace{-.2cm} 0.021  & \hspace{-.2cm} \textbf{0.093}  & \hspace{-.2cm} 0.050  & \hspace{-.2cm} 0.084  & \hspace{-.2cm} 0.082  & \hspace{-.2cm} 0.066  & \hspace{-.2cm} 0.093  & \hspace{-.2cm} 0.045  & \hspace{-.2cm} 0.024  & \hspace{-.2cm} 0.011  & \hspace{-.2cm} 0.072  \\
 & $IMSE(3NB)$& \hspace{-.2cm} 0.103  & \hspace{-.2cm} 0.059  & \hspace{-.2cm} 0.048  & \hspace{-.2cm} 0.023  & \hspace{-.2cm} 0.021  & \hspace{-.2cm} 0.093  & \hspace{-.2cm} 0.050  & \hspace{-.2cm} 0.084  & \hspace{-.2cm} 0.082  & \hspace{-.2cm} 0.066  & \hspace{-.2cm} 0.093  & \hspace{-.2cm} 0.045  & \hspace{-.2cm} \textbf{0.024}  & \hspace{-.2cm} \textbf{\textbf{0.011}}  & \hspace{-.2cm} 0.072  \\
\hline
 \multirow{9}{*}{sym8 \hspace{-.3cm}}
 & $IMSE(1L)$& \hspace{-.2cm} 0.101   & \hspace{-.2cm} 0.050  & \hspace{-.2cm} \textbf{0.024}  & \hspace{-.2cm} 0.012  & \hspace{-.2cm} 0.005  & \hspace{-.2cm} 0.148  & \hspace{-.2cm} 0.072  & \hspace{-.2cm} 0.035  & \hspace{-.2cm} 0.014  & \hspace{-.2cm} \textbf{0.006}  & \hspace{-.2cm} 0.381  & \hspace{-.2cm} 0.178  & \hspace{-.2cm} 0.089  & \hspace{-.2cm} 0.024  & \hspace{-.2cm} 0.012  \\
 & $IMSE(2L)$& \hspace{-.2cm} 0.101   & \hspace{-.2cm} \textbf{0.050}  & \hspace{-.2cm} 0.024  & \hspace{-.2cm} 0.012  & \hspace{-.2cm} 0.005  & \hspace{-.2cm} 0.148  & \hspace{-.2cm} 0.072  & \hspace{-.2cm} 0.035  & \hspace{-.2cm} 0.014  & \hspace{-.2cm} 0.006  & \hspace{-.2cm} 0.381  & \hspace{-.2cm} 0.178  & \hspace{-.2cm} 0.089  & \hspace{-.2cm} 0.024  & \hspace{-.2cm} 0.012  \\
 & $IMSE(3L)$& \hspace{-.2cm} \textbf{0.101}   & \hspace{-.2cm} 0.050  & \hspace{-.2cm} 0.024  & \hspace{-.2cm} 0.012  & \hspace{-.2cm} 0.005  & \hspace{-.2cm} 0.148  & \hspace{-.2cm} 0.072  & \hspace{-.2cm} 0.035  & \hspace{-.2cm} \textbf{0.014}  & \hspace{-.2cm} 0.006  & \hspace{-.2cm} 0.381  & \hspace{-.2cm} 0.178  & \hspace{-.2cm} 0.089  & \hspace{-.2cm} 0.024  & \hspace{-.2cm} 0.012  \\
 & $IMSE(1NT)$& \hspace{-.2cm} 0.139  & \hspace{-.2cm} 0.072  & \hspace{-.2cm} 0.116  & \hspace{-.2cm} 0.022  & \hspace{-.2cm} 0.065  & \hspace{-.2cm} 0.116  & \hspace{-.2cm} 0.080  & \hspace{-.2cm} 0.304  & \hspace{-.2cm} 0.080  & \hspace{-.2cm} 0.133  & \hspace{-.2cm} 0.131  & \hspace{-.2cm} 0.062  & \hspace{-.2cm} 0.465  & \hspace{-.2cm} 0.086  & \hspace{-.2cm} 0.483  \\
 & $IMSE(2NT)$& \hspace{-.2cm} 0.139  & \hspace{-.2cm} 0.072  & \hspace{-.2cm} 0.116  & \hspace{-.2cm} 0.022  & \hspace{-.2cm} 0.065  & \hspace{-.2cm} 0.116  & \hspace{-.2cm} 0.080  & \hspace{-.2cm} 0.304  & \hspace{-.2cm} 0.080  & \hspace{-.2cm} 0.133  & \hspace{-.2cm} 0.131  & \hspace{-.2cm} 0.062  & \hspace{-.2cm} 0.465  & \hspace{-.2cm} 0.086  & \hspace{-.2cm} 0.483  \\
 & $IMSE(3NT)$& \hspace{-.2cm} 0.139  & \hspace{-.2cm} 0.072  & \hspace{-.2cm} 0.116  & \hspace{-.2cm} 0.022  & \hspace{-.2cm} 0.065  & \hspace{-.2cm} 0.116  & \hspace{-.2cm} 0.080  & \hspace{-.2cm} 0.304  & \hspace{-.2cm} 0.080  & \hspace{-.2cm} 0.133  & \hspace{-.2cm} 0.131  & \hspace{-.2cm} 0.063  & \hspace{-.2cm} 0.466  & \hspace{-.2cm} 0.086  & \hspace{-.2cm} 0.483  \\
 & $IMSE(1NB)$& \hspace{-.2cm} 0.104  & \hspace{-.2cm} 0.060  & \hspace{-.2cm} 0.091  & \hspace{-.2cm} 0.025  & \hspace{-.2cm} 0.029  & \hspace{-.2cm} 0.093  & \hspace{-.2cm} 0.044  & \hspace{-.2cm} 0.067  & \hspace{-.2cm} 0.073  & \hspace{-.2cm} 0.076  & \hspace{-.2cm} 0.093  & \hspace{-.2cm} 0.044  & \hspace{-.2cm} 0.041  & \hspace{-.2cm} 0.022  & \hspace{-.2cm} 0.099  \\
 & $IMSE(2NB)$& \hspace{-.2cm} 0.104  & \hspace{-.2cm} 0.060  & \hspace{-.2cm} 0.091  & \hspace{-.2cm} 0.025  & \hspace{-.2cm} 0.029  & \hspace{-.2cm} 0.093  & \hspace{-.2cm} 0.044  & \hspace{-.2cm} 0.067  & \hspace{-.2cm} 0.073  & \hspace{-.2cm} 0.076  & \hspace{-.2cm} 0.093  & \hspace{-.2cm} 0.044  & \hspace{-.2cm} 0.041  & \hspace{-.2cm} 0.022  & \hspace{-.2cm} 0.099  \\
 & $IMSE(3NB)$& \hspace{-.2cm} 0.104  & \hspace{-.2cm} 0.060  & \hspace{-.2cm} 0.091  & \hspace{-.2cm} 0.025  & \hspace{-.2cm} 0.029  & \hspace{-.2cm} 0.093  & \hspace{-.2cm} \textbf{0.044}  & \hspace{-.2cm} 0.067  & \hspace{-.2cm} 0.073  & \hspace{-.2cm} 0.076  & \hspace{-.2cm} \textbf{\textbf{0.093}}  & \hspace{-.2cm} \textbf{0.044}  & \hspace{-.2cm} 0.040  & \hspace{-.2cm} 0.022  & \hspace{-.2cm} 0.099  \\
\hline
\end{tabular}

%% file: Mfresultseqmf1_9999.tex
\begin{tabular}{cc|rrrrr|rrrrr|rrrrr}
\hline
& & \multicolumn{5}{|c|}{SNR=1} & \multicolumn{5}{|c|}{SNR=3} & \multicolumn{5}{|c}{SNR=7} \\
\hline
 & $n$ & 512 & \hspace{-.2cm}1024 & \hspace{-.2cm}2048 & \hspace{-.2cm}4096 & \hspace{-.2cm}8192 & 512 & \hspace{-.2cm}1024 & \hspace{-.2cm}2048 & \hspace{-.2cm}4096 & \hspace{-.2cm}8192 & 512 & \hspace{-.2cm}1024 & \hspace{-.2cm}2048 & \hspace{-.2cm}4096 & \hspace{-.2cm}8192    \\
\hline
 \multirow{9}{*}{db3 \hspace{-.3cm}}
 &IMSE(1L)& \hspace{-.2cm} 14.929  & \hspace{-.2cm} 7.644  & \hspace{-.2cm} 3.727  & \hspace{-.2cm} 1.683  & \hspace{-.2cm} 0.734  & \hspace{-.2cm} 19.993     & \hspace{-.2cm} 10.231     & \hspace{-.2cm} 5.101  & \hspace{-.2cm} 1.938  & \hspace{-.2cm} 0.857  & \hspace{-.2cm} 45.020     & \hspace{-.2cm} 23.056     & \hspace{-.2cm} 12.186     & \hspace{-.2cm} 3.216  & \hspace{-.2cm} 1.530  \\
 &IMSE(2L)& \hspace{-.2cm} 14.929  & \hspace{-.2cm} 7.644  & \hspace{-.2cm} 3.727  & \hspace{-.2cm} 1.683  & \hspace{-.2cm} 0.734  & \hspace{-.2cm} 19.993     & \hspace{-.2cm} 10.231     & \hspace{-.2cm} 5.101  & \hspace{-.2cm} 1.938  & \hspace{-.2cm} 0.857  & \hspace{-.2cm} 45.020     & \hspace{-.2cm} 23.056     & \hspace{-.2cm} 12.186     & \hspace{-.2cm} 3.216  & \hspace{-.2cm} 1.530  \\
 &IMSE(3L)& \hspace{-.2cm} 14.929  & \hspace{-.2cm} 7.644  & \hspace{-.2cm} 3.727  & \hspace{-.2cm} 1.683  & \hspace{-.2cm} 0.734  & \hspace{-.2cm} 19.993     & \hspace{-.2cm} 10.231     & \hspace{-.2cm} 5.101  & \hspace{-.2cm} 1.938  & \hspace{-.2cm} 0.857  & \hspace{-.2cm} 45.020     & \hspace{-.2cm} 23.056     & \hspace{-.2cm} 12.186     & \hspace{-.2cm} 3.216  & \hspace{-.2cm} 1.530  \\
 &IMSE(1NT)& \hspace{-.2cm} 14.349     & \hspace{-.2cm} 7.339  & \hspace{-.2cm} 3.753  & \hspace{-.2cm} 1.870  & \hspace{-.2cm} 1.705  & \hspace{-.2cm} 14.633     & \hspace{-.2cm} 7.359  & \hspace{-.2cm} 3.556  & \hspace{-.2cm} 1.668  & \hspace{-.2cm} 0.811  & \hspace{-.2cm} 15.680     & \hspace{-.2cm} 7.405  & \hspace{-.2cm} 3.471  & \hspace{-.2cm} 1.631  & \hspace{-.2cm} 0.755  \\
 &IMSE(2NT)& \hspace{-.2cm} 14.349     & \hspace{-.2cm} 7.339  & \hspace{-.2cm} 3.753  & \hspace{-.2cm} 1.870  & \hspace{-.2cm} 1.705  & \hspace{-.2cm} 14.633     & \hspace{-.2cm} 7.359  & \hspace{-.2cm} 3.556  & \hspace{-.2cm} 1.668  & \hspace{-.2cm} 0.811  & \hspace{-.2cm} 15.680     & \hspace{-.2cm} 7.405  & \hspace{-.2cm} 3.471  & \hspace{-.2cm} 1.631  & \hspace{-.2cm} 0.755  \\
 &IMSE(3NT)& \hspace{-.2cm} 14.349     & \hspace{-.2cm} 7.339  & \hspace{-.2cm} 3.753  & \hspace{-.2cm} 1.870  & \hspace{-.2cm} 1.705  & \hspace{-.2cm} 14.633     & \hspace{-.2cm} 7.359  & \hspace{-.2cm} 3.556  & \hspace{-.2cm} 1.668  & \hspace{-.2cm} 0.811  & \hspace{-.2cm} 15.680     & \hspace{-.2cm} 7.405  & \hspace{-.2cm} 3.471  & \hspace{-.2cm} 1.631  & \hspace{-.2cm} 0.755  \\
 &IMSE(1NB)& \hspace{-.2cm} 14.303     & \hspace{-.2cm} 7.316  & \hspace{-.2cm} 3.556  & \hspace{-.2cm} 1.652  & \hspace{-.2cm} 0.718  & \hspace{-.2cm} 14.313     & \hspace{-.2cm} 7.305  & \hspace{-.2cm} 3.507  & \hspace{-.2cm} 1.642  & \hspace{-.2cm} 0.710  & \hspace{-.2cm} 13.988     & \hspace{-.2cm} 7.154  & \hspace{-.2cm} 3.441  & \hspace{-.2cm} 1.597  & \hspace{-.2cm} 0.713  \\
 &IMSE(2NB)& \hspace{-.2cm} 14.303     & \hspace{-.2cm} 7.316  & \hspace{-.2cm} 3.556  & \hspace{-.2cm} 1.652  & \hspace{-.2cm} \textbf{0.718}  & \hspace{-.2cm} 14.313     & \hspace{-.2cm} 7.305  & \hspace{-.2cm} 3.507  & \hspace{-.2cm} 1.642  & \hspace{-.2cm} 0.710  & \hspace{-.2cm} 13.988     & \hspace{-.2cm} 7.154  & \hspace{-.2cm} 3.441  & \hspace{-.2cm} 1.597  & \hspace{-.2cm} 0.713  \\
 &IMSE(3NB)& \hspace{-.2cm} 14.303     & \hspace{-.2cm} 7.316  & \hspace{-.2cm} 3.556  & \hspace{-.2cm} 1.652  & \hspace{-.2cm} 0.718  & \hspace{-.2cm} 14.313     & \hspace{-.2cm} 7.305  & \hspace{-.2cm} 3.507  & \hspace{-.2cm} 1.642  & \hspace{-.2cm} 0.710  & \hspace{-.2cm} 13.988     & \hspace{-.2cm} 7.154  & \hspace{-.2cm} 3.441  & \hspace{-.2cm} 1.597  & \hspace{-.2cm} 0.713  \\
\hline
 \multirow{9}{*}{db6 \hspace{-.3cm}}
 &IMSE(1L)& \hspace{-.2cm} 15.078  & \hspace{-.2cm} 7.671  & \hspace{-.2cm} 3.721  & \hspace{-.2cm} 1.679  & \hspace{-.2cm} 0.730  & \hspace{-.2cm} 21.382     & \hspace{-.2cm} 10.488     & \hspace{-.2cm} 5.049  & \hspace{-.2cm} 1.917  & \hspace{-.2cm} 0.841  & \hspace{-.2cm} 52.649     & \hspace{-.2cm} 24.468     & \hspace{-.2cm} 11.914     & \hspace{-.2cm} 3.120  & \hspace{-.2cm} 1.460  \\
 &IMSE(2L)& \hspace{-.2cm} 15.078  & \hspace{-.2cm} 7.671  & \hspace{-.2cm} 3.721  & \hspace{-.2cm} 1.679  & \hspace{-.2cm} 0.730  & \hspace{-.2cm} 21.382     & \hspace{-.2cm} 10.488     & \hspace{-.2cm} 5.049  & \hspace{-.2cm} 1.917  & \hspace{-.2cm} 0.841  & \hspace{-.2cm} 52.649     & \hspace{-.2cm} 24.468     & \hspace{-.2cm} 11.914     & \hspace{-.2cm} 3.120  & \hspace{-.2cm} 1.460  \\
 &IMSE(3L)& \hspace{-.2cm} 15.078  & \hspace{-.2cm} 7.671  & \hspace{-.2cm} 3.721  & \hspace{-.2cm} 1.679  & \hspace{-.2cm} 0.730  & \hspace{-.2cm} 21.382     & \hspace{-.2cm} 10.488     & \hspace{-.2cm} 5.049  & \hspace{-.2cm} 1.917  & \hspace{-.2cm} 0.841  & \hspace{-.2cm} 52.649     & \hspace{-.2cm} 24.468     & \hspace{-.2cm} 11.914     & \hspace{-.2cm} 3.120  & \hspace{-.2cm} 1.460  \\
 &IMSE(1NT)& \hspace{-.2cm} 14.366     & \hspace{-.2cm} 7.326  & \hspace{-.2cm} 3.700  & \hspace{-.2cm} 1.793  & \hspace{-.2cm} 0.910  & \hspace{-.2cm} 14.849     & \hspace{-.2cm} 7.337  & \hspace{-.2cm} 3.548  & \hspace{-.2cm} 1.681  & \hspace{-.2cm} 0.797  & \hspace{-.2cm} 16.904     & \hspace{-.2cm} 7.273  & \hspace{-.2cm} 3.475  & \hspace{-.2cm} 1.634  & \hspace{-.2cm} 0.737  \\
 &IMSE(2NT)& \hspace{-.2cm} 14.366     & \hspace{-.2cm} 7.326  & \hspace{-.2cm} 3.700  & \hspace{-.2cm} 1.793  & \hspace{-.2cm} 0.910  & \hspace{-.2cm} 14.849     & \hspace{-.2cm} 7.337  & \hspace{-.2cm} 3.548  & \hspace{-.2cm} 1.681  & \hspace{-.2cm} 0.797  & \hspace{-.2cm} 16.904     & \hspace{-.2cm} 7.273  & \hspace{-.2cm} 3.475  & \hspace{-.2cm} 1.634  & \hspace{-.2cm} 0.737  \\
 &IMSE(3NT)& \hspace{-.2cm} 14.366     & \hspace{-.2cm} 7.326  & \hspace{-.2cm} 3.701  & \hspace{-.2cm} 1.793  & \hspace{-.2cm} 0.910  & \hspace{-.2cm} 14.849     & \hspace{-.2cm} 7.337  & \hspace{-.2cm} 3.548  & \hspace{-.2cm} 1.681  & \hspace{-.2cm} 0.797  & \hspace{-.2cm} 16.904     & \hspace{-.2cm} 7.273  & \hspace{-.2cm} 3.475  & \hspace{-.2cm} 1.634  & \hspace{-.2cm} 0.737  \\
 &IMSE(1NB)& \hspace{-.2cm} 14.302     & \hspace{-.2cm} \textbf{7.315}  & \hspace{-.2cm} 3.557  & \hspace{-.2cm} 1.650  & \hspace{-.2cm} 0.746  & \hspace{-.2cm} 14.312     & \hspace{-.2cm} 7.305  & \hspace{-.2cm} 3.507  & \hspace{-.2cm} 1.642  & \hspace{-.2cm} 0.708  & \hspace{-.2cm} \textbf{13.987}     & \hspace{-.2cm} 7.153  & \hspace{-.2cm} 3.441  & \hspace{-.2cm} 1.596  & \hspace{-.2cm} 0.713  \\
 &IMSE(2NB)& \hspace{-.2cm} 14.302     & \hspace{-.2cm} 7.315  & \hspace{-.2cm} 3.557  & \hspace{-.2cm} 1.650  & \hspace{-.2cm} 0.746  & \hspace{-.2cm} 14.312     & \hspace{-.2cm} 7.305  & \hspace{-.2cm} 3.507  & \hspace{-.2cm} 1.642  & \hspace{-.2cm} 0.708  & \hspace{-.2cm} 13.987     & \hspace{-.2cm} 7.153  & \hspace{-.2cm} 3.441  & \hspace{-.2cm} 1.596  & \hspace{-.2cm} 0.713  \\
 &IMSE(3NB)& \hspace{-.2cm} 14.302     & \hspace{-.2cm} 7.315  & \hspace{-.2cm} 3.557  & \hspace{-.2cm} \textbf{1.650}  & \hspace{-.2cm} 0.746  & \hspace{-.2cm} \textbf{14.312}     & \hspace{-.2cm} 7.305  & \hspace{-.2cm} 3.507  & \hspace{-.2cm} 1.642  & \hspace{-.2cm} 0.708  & \hspace{-.2cm} 13.987     & \hspace{-.2cm} 7.153  & \hspace{-.2cm} \textbf{3.441}  & \hspace{-.2cm} 1.596  & \hspace{-.2cm} 0.713  \\
\hline
 \multirow{9}{*}{sym8 \hspace{-.3cm}}
 &IMSE(1L)& \hspace{-.2cm} 14.874  & \hspace{-.2cm} 7.591  & \hspace{-.2cm} 3.685  & \hspace{-.2cm} 1.676  & \hspace{-.2cm} 0.729  & \hspace{-.2cm} 19.526     & \hspace{-.2cm} 9.767  & \hspace{-.2cm} 4.717  & \hspace{-.2cm} 1.882  & \hspace{-.2cm} 0.823  & \hspace{-.2cm} 42.498     & \hspace{-.2cm} 20.545     & \hspace{-.2cm} 10.086     & \hspace{-.2cm} 2.922  & \hspace{-.2cm} 1.363  \\
 &IMSE(2L)& \hspace{-.2cm} 14.874  & \hspace{-.2cm} 7.591  & \hspace{-.2cm} 3.685  & \hspace{-.2cm} 1.676  & \hspace{-.2cm} 0.729  & \hspace{-.2cm} 19.526     & \hspace{-.2cm} 9.767  & \hspace{-.2cm} 4.717  & \hspace{-.2cm} 1.882  & \hspace{-.2cm} 0.823  & \hspace{-.2cm} 42.498     & \hspace{-.2cm} 20.545     & \hspace{-.2cm} 10.086     & \hspace{-.2cm} 2.922  & \hspace{-.2cm} 1.363  \\
 &IMSE(3L)& \hspace{-.2cm} 14.874  & \hspace{-.2cm} 7.591  & \hspace{-.2cm} 3.685  & \hspace{-.2cm} 1.676  & \hspace{-.2cm} 0.729  & \hspace{-.2cm} 19.526     & \hspace{-.2cm} 9.767  & \hspace{-.2cm} 4.717  & \hspace{-.2cm} 1.882  & \hspace{-.2cm} 0.823  & \hspace{-.2cm} 42.498     & \hspace{-.2cm} 20.545     & \hspace{-.2cm} 10.086     & \hspace{-.2cm} 2.922  & \hspace{-.2cm} 1.363  \\
 &IMSE(1NT)& \hspace{-.2cm} 14.365     & \hspace{-.2cm} 7.329  & \hspace{-.2cm} 3.660  & \hspace{-.2cm} 1.798  & \hspace{-.2cm} 1.353  & \hspace{-.2cm} 14.845     & \hspace{-.2cm} 7.332  & \hspace{-.2cm} 3.535  & \hspace{-.2cm} 1.705  & \hspace{-.2cm} 0.794  & \hspace{-.2cm} 16.892     & \hspace{-.2cm} 7.290  & \hspace{-.2cm} 3.461  & \hspace{-.2cm} 1.627  & \hspace{-.2cm} 0.757  \\
 &IMSE(2NT)& \hspace{-.2cm} 14.365     & \hspace{-.2cm} 7.329  & \hspace{-.2cm} 3.660  & \hspace{-.2cm} 1.798  & \hspace{-.2cm} 1.353  & \hspace{-.2cm} 14.845     & \hspace{-.2cm} 7.332  & \hspace{-.2cm} 3.535  & \hspace{-.2cm} 1.705  & \hspace{-.2cm} 0.794  & \hspace{-.2cm} 16.892     & \hspace{-.2cm} 7.290  & \hspace{-.2cm} 3.461  & \hspace{-.2cm} 1.627  & \hspace{-.2cm} 0.757  \\
 &IMSE(3NT)& \hspace{-.2cm} 14.365     & \hspace{-.2cm} 7.329  & \hspace{-.2cm} 3.660  & \hspace{-.2cm} 1.798  & \hspace{-.2cm} 1.353  & \hspace{-.2cm} 14.845     & \hspace{-.2cm} 7.332  & \hspace{-.2cm} 3.535  & \hspace{-.2cm} 1.705  & \hspace{-.2cm} 0.794  & \hspace{-.2cm} 16.892     & \hspace{-.2cm} 7.290  & \hspace{-.2cm} 3.461  & \hspace{-.2cm} 1.627  & \hspace{-.2cm} 0.757  \\
 &IMSE(1NB)& \hspace{-.2cm} 14.302     & \hspace{-.2cm} 7.315  & \hspace{-.2cm} \textbf{3.552}  & \hspace{-.2cm} 1.673  & \hspace{-.2cm} 0.718  & \hspace{-.2cm} 14.313     & \hspace{-.2cm} \textbf{7.304}  & \hspace{-.2cm} 3.507  & \hspace{-.2cm} 1.642  & \hspace{-.2cm} 0.708  & \hspace{-.2cm} 13.987     & \hspace{-.2cm} 7.153  & \hspace{-.2cm} 3.441  & \hspace{-.2cm} \textbf{1.596}  & \hspace{-.2cm} 0.713  \\
 &IMSE(2NB)& \hspace{-.2cm} 14.302     & \hspace{-.2cm} 7.315  & \hspace{-.2cm} 3.552  & \hspace{-.2cm} 1.673  & \hspace{-.2cm} 0.718  & \hspace{-.2cm} 14.313     & \hspace{-.2cm} 7.304  & \hspace{-.2cm} 3.507  & \hspace{-.2cm} 1.642  & \hspace{-.2cm} \textbf{0.708}  & \hspace{-.2cm} 13.987     & \hspace{-.2cm} 7.153  & \hspace{-.2cm} 3.441  & \hspace{-.2cm} 1.596  & \hspace{-.2cm} 0.713  \\
 &IMSE(3NB)& \hspace{-.2cm} \textbf{14.302}     & \hspace{-.2cm} 7.315  & \hspace{-.2cm} 3.552  & \hspace{-.2cm} 1.673  & \hspace{-.2cm} 0.718  & \hspace{-.2cm} 14.313     & \hspace{-.2cm} 7.304  & \hspace{-.2cm} \textbf{3.507}  & \hspace{-.2cm} \textbf{1.642}  & \hspace{-.2cm} 0.708  & \hspace{-.2cm} 13.987     & \hspace{-.2cm} \textbf{7.153}  & \hspace{-.2cm} 3.441  & \hspace{-.2cm} 1.596  & \hspace{-.2cm} \textbf{0.713}  \\
\hline
\end{tabular}